\documentclass[aps,prd,twocolumn,groupedaddress,nofootinbib]{revtex4-2}

\usepackage{mathtools}
\usepackage{amsmath}   
\usepackage{amssymb}
\usepackage{graphicx}
\usepackage{xcolor}
\usepackage{multirow}
\usepackage{soul}
\usepackage{aas_macros} 

\usepackage{soul}

\newcommand{\pder}[2]{\frac{\partial #1}{\partial #2}}

\newcommand{\vect}[1]{\boldsymbol{\mathbf{#1}}}

\newcommand{\E}[1]{\times 10^{#1}}

\newcommand{\Msun}{M_\odot}

\newcommand{\orcid}[1]{\href{https://orcid.org/#1}{\includegraphics[width=9pt]{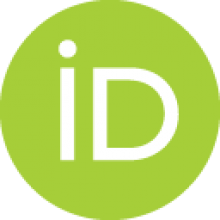}}}

\begin{document}

\title[Collapsar Disk Outflows]{Collapsar disk outflows I: Viscous hydrodynamic evolution in axisymmetry}

\author{Coleman Dean\orcid{0000-0001-9364-4785}}
\email[]{cddean@ualberta.ca}
\affiliation{Department of Physics, University of Alberta, Edmonton, AB T6G 2E1, Canada}

\author{Rodrigo Fern\'andez\orcid{0000-0003-4619-339X}}
\affiliation{Department of Physics, University of Alberta, Edmonton, AB T6G 2E1, Canada}

\date{\today}

\begin{abstract}
We investigate mass ejection from accretion disks formed during the collapse of rapidly-rotating Wolf-Rayet stars. The neutrino-cooled, 
black hole (BH) accretion disk system that forms at the center of the star -- and the ensuing outflows -- provide the conditions for 
these systems to be candidate $r$-process element production sites and potential progenitors of broad-lined Type Ic (Ic-BL) supernovae. 
Here we present global, long-term axisymmetric hydrodynamic simulations of collapsar disks that include angular momentum transport 
through shear viscosity, neutrino emission and absorption, a  
19-isotope nuclear reaction network and nuclear statistical 
equilibrium solver, a pseudo-Newtonian BH with mass and spin modified by accreted matter, and self-gravity. Starting from a stellar 
profile collapsed in spherical symmetry, our models capture disk formation self-consistently, and are evolved until after 
the shock wave -- driven by disk winds -- reaches the surface of the star. None of our models achieve sufficient 
neutronization to eject significant amounts of $r$-process elements (detailed nucleosynthesis calculations will follow in 
a companion paper). Sufficient $^{56}$Ni is produced to power a typical type Ic-BL supernova light curve, but the average asymptotic velocity is a 
factor $\sim 2-3$ times too slow to account for the typical line widths in type Ic-BL supernova spectra. The gap in neutrino 
emission between BH formation and shocked disk formation, and the magnitude of the subsequent peak in emission, would be observable diagnostics 
of the internal conditions of the progenitor in a galactic collapsar. Periodic oscillations of the shocked disk prior to its expansion are also 
a potential observable through their impact on the the neutrino and gravitational wave signals.
\end{abstract}

\maketitle


\section{Introduction \label{sec:intro}}

The detection of numerous black hole (BH) binary
mergers by the LIGO-Virgo Collaboration (\cite{GWTC1,GWTC2,GWTC2.1,GWTC3}) has increased interest
in the origin of stellar-mass BHs. With transient surveys expanding
the known parameter space of time-domain
astronomy, explosive stellar events such as supernovae 
(SNe) have been found to show diversity beyond established classes 
(e.g., \cite{mm_2018,graham_2019,modjaz_2019}). 
Progress in our understanding of the formation of stellar mass  
BHs thus requires theoretical characterization of the associated electromagnetic (EM) signatures 
of these events, to maximize the insight gained from observations. 

The core-collapse of massive stars is thought to be the dominant 
formation path for stellar-mass BHs. When the progenitor mass is below the limit for the
onset of pair instability, collapse always leads to the formation of a protoneutron star \cite{oconnor_2011}, 
with subsequent failure of the SN (e.g., \cite{nadyozhin_1980}), or fallback accretion in an otherwise successful 
SN (e.g. \cite{colgate_1971}), leading to BH formation. Very massive stars ($M \gtrsim 250\,M_\odot$)
can also lead directly to BH formation \cite{fryer_2001}. 

The \emph{collapsar} model \cite{woosley_1993} describes a massive progenitor star with
significant rotation at the time of core collapse, which fails to explode as a
standard SN and forms a central BH. Collapsing material circularizes outside the innermost
stable circular orbit (ISCO), forming an accretion disk. The location of disk formation depends crucially 
on the angular momentum profile of the progenitor, which generally is not well-known for massive stars.
If the disk forms close enough to the BH for neutrino cooling to
become important, a relativistic jet can be launched, resulting in a long gamma-ray burst (GRB) (e.g., 
\cite{macfadyen_1999}). An associated SN explosion could be powered by accretion disk winds
\cite{macfadyen_2003}, or via a relativistic jet cocoon that shocks and unbinds
the star (e.g. \cite{macfadyen_2001,gottlieb_2022}). If the
circularization radius is too large for neutrino cooling to be important, 
an explosion that ejects the outer stellar layers can still be produced, but likely 
with a lower 
energy than standard SNe (e.g., \cite{bodenheimer_1983,antoni_2023}). 

Collapsars have been proposed as a site of rapid neutron capture ($r$-process) element production (\cite{macfadyen_1999,kohri_2005}),  having a shorter delay timescale after star formation  
than neutron star (NS) mergers  (e.g., \cite{siegel_2019}), which need to experience orbital 
decay by gravitational wave emission before merging \cite{peters_1963}. 
The neutron-rich conditions for the $r$-process occur when the collapsar disk achieves high enough densities that electrons 
are degenerate, and neutrino interactions are important (e.g., \cite{beloborodov_2003,chen_2007}).

Production of $r$-process elements 
with short delay timescales may be needed to explain the europium abundances in low metallicity stars in dwarf 
galaxies within the local group, \cite{ji_2016} as well as the evolution of the ratio of europium to iron in our own galaxy
(e.g., \cite{cote_2017,hotokezaka_2018,zevin_2019,kobayashi_2023}).
Whether collapsars have indeed the ability to contribute with significant amounts of 
$r$-process elements
remains an open question, however, as the neutron-rich matter must be ejected from the system. 
Recent evidence in favor of this hypothesis 
is the claimed detection of a kilonova from a long GRB  \cite{rastinejad_2022}. 

Here we study the long-term evolution of collapsar disks and their outflows using two dimensional (2D) 
viscous hydrodynamic simulations that include neutrino emission and absorption, as well as nuclear energy 
release. While our simulations are Newtonian, the BH is treated using a spinning pseudo-Newtonian potential,
which allows for a good estimate of sub-relativistic, accretion-powered 
mass ejection at large radii (we cannot obtain a jet 
and investigate the production of a long GRB and/or a cocoon-driven explosion, however). 
The initial condition is obtained by evolving a rotating progenitor with a spherically-symmetric, general 
relativistic neutrino radiation-hydrodynamic code until BH formation. 
Our disk simulations explore variations
in the strength of viscous angular momentum transport, as well as
in progenitor stars, and in equation of state (EOS) used prior to 
BH formation. 
This paper focuses on the disk evolution and mass ejection, a companion paper will investigate 
the detailed nucleosynthesis signatures of the disk outflow.

The structure of the paper is the following. Section \ref{sec:methods} describes our choice of progenitor 
stars, evolution up to the point of BH formation, physical assumptions and numerical setup for
axisymmetric simulations, choice of model parameters, and analysis methods. 
The results are discussed in Section \ref{sec:results}, including an overview of disk evolution, 
properties of the disk outflow, neutronization and neutrino emission, potential to power broad-line 
type Ic (Ic-BL) SNe, 
and a comparison of our results to similar work by other groups. A summary and discussion follow in Section \ref{sec:summary}.
The appendices describe our implementation of nuclear burning and nuclear statistical equilibrium,
and the floors used in axisymmetric simulations.


\section{Methods \label{sec:methods}}

\subsection{Progenitors and Evolution to BH Formation \label{sec:progenitor}}

\begin{figure*}
\includegraphics[width=\textwidth]{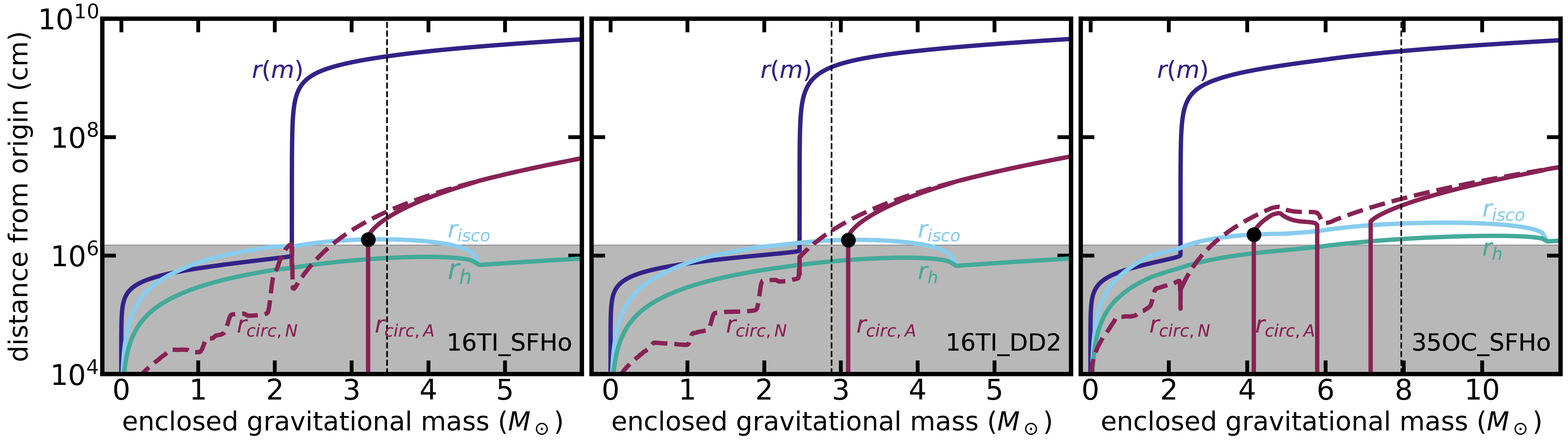}
\caption{Characteristic radii as a function of enclosed gravitational mass
at the last snapshot before BH formation in \texttt{GR1D} (Section \ref{sec:progenitor}) for the
presupernova progenitors \texttt{16TI} (SFHo EOS left, DD2 EOS
center) and \texttt{35OC} (SFHo EOS right),
evolved in all cases with approximate rotation effects. Curves show the radial 
coordinate (purple), ISCO radius (light blue), event horizon radius (green), and
circularization radii obtained with a Newtonian potential (dashed burgundy,
eq.~\ref{eqn:rcirc_newtonian}) and with the Artemova pseudo-Newtonian potential
(solid burgundy, defined by Equation~\ref{eqn:jkepler_artemova}). 
All quantities account for the spinup of the BH with the
enclosed angular momentum at each mass (eq.~\ref{eqn:bh_spin}). The gray shaded
area shows the region excised initially from the computational domain for subsequent
evolution in 2D with \texttt{FLASH} (Section~\ref{sec:flash}). The black circle marks the predicted BH mass at
disk formation, and the vertical dashed black line shows the 
actual BH mass when the shocked accretion disk forms in the \texttt{FLASH} simulation.}
\label{fig:radr_mgrv}
\end{figure*}

We employ two stellar progenitors from \cite{woosley_2006b}
which undergo chemically homogeneous evolution and reach the presupernova state
as Wolf-Rayet stars. Model \texttt{16TI} is a $16\,M_\odot$ zero age main sequence (ZAMS) star with metallicity $1\%$ solar
and presupernova mass $14\,M_\odot$, and model \texttt{35OC} is a $35\,M_\odot$ ZAMS star with
metallicity $10\%$ solar and presupernova mass $28\,M_\odot$. Both are evolved
including a prescription for magnetic torques and reduced mass loss rates, and
have previously been used in global collapsar simulations (e.g.,
\cite{harikae_2009a,lopezcamara_2009,lindner_2010,obergaulinger_2017,just_2022}).

Progenitors are evolved until BH formation with the spherically-symmetric,
neutrino radiation-hydrodynamic code {\tt GR1D} version 1 \cite{oconnor_2010}.
The code solves the equations of general-relativistic hydrodynamics with a
finite-volume method, and employs a three-flavor gray leakage scheme to treat
neutrino emission and absorption. Our default evolution mode employs the
SFHo EOS \cite{steiner_2013}, with one model using the DD2 EOS \cite{hempel_2012} 
to quantify sensitivity to BH formation
time.  The computational grid is uniform inside $20$\,km, and expands
logarithmically outside until a radius $\sim 10^9$\,cm at which the density is
$2\times 10^3$\,g\,cm$^{-3}$, with a total resolution of $1000-1200$ cells
depending on progenitor. BH formation is deemed to have occurred when the
central density increases rapidly with time toward $\gtrsim 10^{15}$\,g\,cm$^{-3}$,
accompanied by a rapid decrease of the central value of the lapse function
toward zero, at which point the code crashes. See \cite{ivanov_2021} for
more details about simulation parameters and verification tests.

All cases are evolved accounting for rotation in \texttt{GR1D}, starting from the initial
angular momentum distribution of the star. \texttt{GR1D} includes an approximate
prescription for angle-averaged rotation that accounts for centrifugal
effects and conservation of angular momentum \cite{oconnor_2010}. 
While this approximation provides a reasonable estimate to the delay
until BH formation due to rotation effects, it cannot capture multi-dimensional
phenomena such as the formation of transient accretion disks during the protoneutron
star phase (e.g., the \texttt{16TI} progenitor evolution in \cite{obergaulinger_2022}).

Figure~\ref{fig:radr_mgrv} shows spatial profiles at the last snapshot before BH
formation in \texttt{GR1D}.  The sharp increase in the radial coordinate with enclosed mass
occurs at the surface of the protoneutron star. Also shown are the ISCO and horizon radii of
a BH of mass equal to the enclosed gravitational mass and dimensionless spin
$a_{\rm bh}$ implied by the enclosed angular momentum
\begin{equation}
\label{eqn:bh_spin}
a_{\rm bh} = \frac{J_{\rm bh}/M_{\rm bh}}{r_{\rm g}\,c} = \frac{c\, J_{\rm bh}}{GM_{\rm bh}^2},
\end{equation}
where $J_{\rm bh}$ is the total angular momentum and $M_{\rm bh}$ the
gravitational mass of a BH that would form at that mass coordinate
(as usual, $r_{\rm g}\equiv GM_{\rm bh}/c^2$). The ISCO and horizon
radii are computed using the analytic formulae for the Kerr metric (e.g.,
\cite{bardeen_1972}), while the total angular momentum enclosed at each mass
coordinate is computed consistently with the coordinate system in \texttt{GR1D}
(equation~16 of \cite{oconnor_2010}).

As the star continues to collapse, the BH grows in mass and changes its spin
by accreting matter, sweeping through the Lagrangian mass coordinate
in Figure~\ref{fig:radr_mgrv}. The subsequent evolution of collapsars is normally
characterized by the Newtonian circularization radius
\begin{equation}
\label{eqn:rcirc_newtonian}
r_{\rm circ,N} = \frac{j^2}{GM_{\rm g}},
\end{equation}
where $j(M_{\rm g})$ is the specific angular momentum and $M_{\rm g}$ is the
enclosed gravitational mass. At this location, the centrifugal acceleration
balances the Newtonian acceleration of gravity at the equator. The circularization radius increases outward
because the specific angular momentum in these progenitors
increases faster than the square root of the enclosed gravitational mass (c.f., Figure 2 of \cite{woosley_2006b}).

Our post-BH evolution (Section \ref{sec:flash}) employs a pseudo-Newtonian potential $\Phi_{\rm bh}$
to model the gravity of the BH, which yields a circularization radius $r_{\rm circ,A}$
that differs from the Newtonian value in Equation~(\ref{eqn:rcirc_newtonian}). We use the potential of \cite{artemova1996},
which provides an ISCO for a spinning BH \cite{FKMQ14}:
\begin{equation}
\label{eqn:artemova_potential}
\Phi_{\rm bh}(r) =
\begin{dcases}
\frac{GM_{\rm bh}}{(\beta-1) r_{\rm h}}\left[1 - \left(\frac{r}{r-r_{\rm h}} \right)^{\beta-1} \right]
& \qquad (\beta\ne 1)\\
\noalign{\smallskip}
\frac{GM_{\rm bh}}{r_{\rm h}}\ln\left(1- \frac{r_{\rm h}}{r}\right)
& \qquad (\beta=1)
\end{dcases}
\end{equation}
where $r_{\rm h}$ is the horizon radius, and
\begin{equation}
\beta = \frac{r_{\rm isco}}{r_{\rm h}} - 1
\end{equation}
with $r_{\rm isco}$ the ISCO radius. In the absence of spin, $\beta = 2$ and the potential
is identical to that of \cite{paczynsky1980}. For arbitrary spins, $r_{\rm isco}$ and
$r_{\rm h}$ are computed analytically as in the Kerr metric \cite{bardeen_1972}.
The Keplerian specific angular momentum in the 
potential of Equation~(\ref{eqn:artemova_potential}) can be obtained by balancing the
gravitational and centrifugal accelerations at the equator
\begin{equation}
\label{eqn:jkepler_artemova}
j^2_{\rm K} = GM_{\rm bh}r\left(1-r_{\rm h}/r \right)^{-\beta}.
\end{equation}
For a given specific angular momentum $j$ and enclosed gravitational mass $M_{\rm g}$ in the progenitor, 
inverting equation~(\ref{eqn:jkepler_artemova}) for $r$, 
setting $j_{\rm K}=j$ and $M_{\rm bh}=M_{\rm g}$, yields the circularization 
radius $r_{\rm circ,A}$ shown in Figure~\ref{fig:radr_mgrv} for the progenitors we 
consider in this study. The resulting value is equal or smaller than
the Newtonian circularization radius, and at small specific angular momenta 
there is no solution.

When $r_{\rm circ,A} \gtrsim r_{\rm isco}$, a 
shocked accretion disk is expected to form. Thereafter, accretion of matter
with higher angular momentum should be halted, and the characteristic radii in
Figure~\ref{fig:radr_mgrv} are no longer predictive for higher enclosed masses.
This includes the point where $r_{\rm isco}$ and $r_{\rm h}$ merge at high
enclosed mass, which would occur if the BH achieved maximal rotation, but
does not occur in practice due to the existence of the accretion disk.

\subsection{Evolution after BH formation \label{sec:flash}}

Once a BH forms in \texttt{GR1D}, we use the spatial distribution
of thermodynamic and kinematic quantities as initial conditions for subsequent
evolution, which we carry out in two-dimensional (2D) axisymmetry using \texttt{FLASH}. The mapping procedure
is similar to that reported in \cite{ivanov_2021}, using pressure, density,
and composition as inputs to the EOS in order to minimize transients. 
The specific angular momentum profile 
from \texttt{GR1D} is mapped assuming 
cylindrical symmetry, i.e. $j(r,\theta) \propto \sin^2(\theta)$. 

We use {\tt FLASH} version 3.2 \cite{fryxell00,dubey2009}
to solve the equations of mass, momentum, energy, and baryon/lepton/charge conservation
in 2D axisymmetric spherical coordinates $(r,\theta)$, with source terms due to
gravity, shear viscosity, neutrino emission/absorption, and nuclear
reactions 
\begin{eqnarray}
\label{eqn:mass_conservation}
\frac{\partial \rho}{\partial t} + \nabla \cdot (\rho\mathbf{v_{\rm p}}) & = & 0 \\
\label{eqn:momentum_conservation}
\frac{D \mathbf{v_{\rm p}}}{D t} & = & - \frac{\Delta P}{\rho}
-\nabla\Phi \\
\label{eqn:angular_conservation}
\rho\frac{D j}{D t} & = & r\sin\theta\,(\nabla\cdot T)_\phi \\ 
\label{eqn:energy_conservation}
\rho\frac{D \epsilon}{D t} + P\nabla\cdot\mathbf{v_{\rm p}}
& = & \frac{1}{\rho\nu}T:T + \rho\left(q_{\rm nuc} + q_{\nu}\right) \\
\label{eqn:poisson}
\nabla^2\Phi & = & 4\pi G \rho + \nabla^2\Phi_{\rm bh} \\
\label{eqn:fuel_evolution}
\frac{\partial \mathbf X}{\partial t} & = & \Theta(\rho,T,\mathbf{X}) + {\Gamma}_{\nu}
\end{eqnarray}
where $D/Dt \equiv \partial/\partial t + \mathbf{v_{\rm p}}\cdot \nabla$, $\mathbf{v_{\rm p}} =
v_r \hat{r} + v_\theta \hat{\theta}$ is the two dimensional (poloidal) velocity, $\rho$ is
the density, $P$ is the pressure, $\epsilon$ is the specific internal energy,
$j$ is the specific angular momentum scalar, $\Phi$ is the gravitational
potential, $T$ is the viscous stress tensor, and $\mathbf{X}$ are the
mass fractions of species considered. The rate of change of the mass fractions
caused by the nuclear network is denoted by $\Theta$, and the specific nuclear heating 
from the network is denoted by $q_{\rm nuc}$. The rate of change of mass
fractions caused by charged-current weak interactions mediated by neutrino
emission and absorption is denoted by $\Gamma_\nu$, and the
specific net neutrino heating rate is denoted by $q_\nu$

We employ the (Helmholtz) equation of state of \cite{timmes2000}, and extend the
tabulated electron-positron quantities for $\rho > 10^{11}$\,g\,cm$^{-3}$ and $T>10^{11}$\,K
with analytic expressions for a relativistic electron-positron gas of arbitrary
degeneracy \cite{bethe80}. At densities below the minimum of the table
($\rho < 10^{-10}$\,g\,cm$^{-3}$) we use an ideal gas law for electrons.
For $T < 5\times 10^9$\,K, we use the 19-isotope nuclear reaction
network of \cite{weaver1978} with the MA28 sparse matrix solver and Bader-Deuflhard variable time
stepping method (e.g., \cite{Timmes1999}). For $T \geq 5\times 10^9$\,K, we set
the abundances of these isotopes to their values in nuclear statistical equilibrim (NSE, Appendix~\ref{sec:NSE}).

The internal energy update (described in Appendix \ref{sec:internal_energy_update}) 
accounts for viscous heating, neutrino heating, and nuclear heating in two separate half-timesteps. The NSE transition temperature is
set initially at $1.4\E{10}$ K for numerical reasons, up until the point of
shock formation, where infalling material begins to form an accretion disk.
Prior to this time, material with sufficiently high temperatures is plunging into
the black hole supersonically. From disk formation onward, the NSE transition temperature is set
to its default value at $5\E{9}$ K. For numerical reasons, NSE is not imposed on fluid within a
factor 10 of the density floor, or for atmospheric material.

Angular momentum transport is included via a shear stress tensor $T$ with
non-zero components $r\phi$ and $\theta\phi$, thus modeling conversion of shear kinetic energy
into heat and turbulence (e.g., \cite{stone1999}). The viscosity coefficient is parameterized
as in \cite{shakura1973}
\begin{equation}
\nu  = \alpha \frac{P/\rho}{\Omega_{\rm K}}
\end{equation}
with the local Keplerian angular frequency defined as
\begin{equation}
\Omega^2_{K} = \frac{1}{r}\frac{d\Phi}{dr}
\end{equation}
(see \cite{FM13} for details). The tensor $T$ modifies $j$ (equation~\ref{eqn:angular_conservation})
and contributes  with a heating term in the energy equation (\ref{eqn:energy_conservation}).
Results of axisymmetric hydrodynamic simulations using this prescription compare favorably with general
relativistic magnetohydrodynamic (MHD) simulations in the advective state \cite{fernandez2019}.
To avoid numerical problems in regions of mostly radial infall, viscous heating and angular momentum
transport are suppressed as $e^{-|\mathbf{v}|/v_\phi}$ for $|\mathbf{v}| > v_\phi \equiv j/(r\sin\theta)$,
following a similar prescription from \cite{macfadyen_1999}. 
We also cap the viscous heating at $10^6$
times the internal energy per timestep, to eliminate numerical issues near the low density
polar funnel. This effectively sets a minimum value for the cooling timestep limiter. 

\begin{table*}
\caption{List of evolved models, parameters used, and key simulation timescales. Columns from left to right show the model name, progenitor star from \cite{woosley_2006b}, EOS used in \texttt{GR1D} evolution to BH formation, and viscosity parameter used in the 2D post-BH evolution. Subsequent columns show times relative to the bounce time in \texttt{GR1D}: BH formation time in \texttt{GR1D}, shocked disk formation time, mass of the BH at disk formation, shock breakout time (leading edge  reaching the surface of the star), and the maximum simulation 
time. \label{tab:progenitors}}
\begin{ruledtabular}
\begin{tabular}{lcccccccc}
Model & Progenitor & EOS & $\alpha$ & $t_{\rm bh}$ (s) & $t_{\rm df}$ (s) & $M_{\rm bh}(t_{\rm df})$ (M$_\odot$) & $t_{\rm sb}$ (s) & $t_{\rm max}$ (s) \\ 
\noalign{\smallskip}
\hline
\noalign{\smallskip}
\texttt{16TI\_SFHo}              & 16TI & SFHo & 0.03 & 2.72 & 11.1 & 3.5 & 116 & 219.8 \\ 
\noalign{\smallskip}
\texttt{16TI\_SFHo\_$\alpha$01}  &      &      & 0.1  & 2.72 & 11.0 & 3.5 & 236 & 427.1 \\
\texttt{16TI\_SFHo\_$\alpha$001} &      &      & 0.01 & 2.72 & 10.6 & 3.4 & 153 & 295.4 \\
\texttt{16TI\_DD2}               &      & DD2  & 0.03 & 5.24 & 9.9  & 2.9 & 168 & 302.0 \\
\texttt{35OC\_SFHo}              & 35OC & SFHo &      & 0.99 & 10.8 & 7.9 & 68  & 102.8 \\
\end{tabular}
\end{ruledtabular}
\end{table*}

We include neutrino emission and absorption via a 3-species leakage scheme for emission and a lighbulb-type approximation for absorption \cite{FM13,MF14,lippuner2017,fernandez_2022}. Emission processes include
electron/positron capture on nucleons, using the rates of \cite{bruenn85}, as well as 
electron-positron pair annihilation and plasmon decay using the rates of \cite{ruffert_1996}. 
Opacities account for charged-current absorption and neutral-current scattering on nucleons.
Emissivities and opacities match those used in the leakage scheme of
{\tt GR1D} for evolution prior to BH formation \cite{oconnor_2010}, with the main
differences between codes being the procedure to
compute  the optical depth and the prescription for absorption.
These neutrino processes contribute with a heating/cooling source term $q_\nu$ in the energy 
equation (Equation \ref{eqn:energy_conservation}), and a rate of change of the mass fractions of 
neutrons and protons ${\Gamma}_\nu$ in the evolution
equation for mass fractions (Equation \ref{eqn:fuel_evolution}).
The electron
fraction is computed from the mass fractions of ions using charge conservation
\begin{equation}
Y_e = \frac{\bar Z}{\bar A},
\end{equation}
with $\bar A = (\sum_i X_i/A_i)^{-1}$ and $\bar Z = \bar A \sum_i (X_i Z_i/A_i)$. Changes in $Y_e$
thus occur implicitly through equation~(\ref{eqn:fuel_evolution}). Additional energy loss
channels that do not alter the composition 
are included in the nuclear reaction network using the analytic fits of \cite{itoh1996},
with an additional correction factor $e^{-\rho_{\rm 11}}$ ($\rho_{\rm 11} = \rho/[10^{11}$\,g\,cm$^{-3}]$)
to account for neutrino trapping in high-density regions.

The Poisson equation (\ref{eqn:poisson}) for the gravitational potential generated by the fluid in the
computational domain is solved with the multipole method of \cite{MuellerSteinmetz1995}, as implemented in \cite{FMM19}.
The BH contribution $\Phi_{\rm bh}$ (Equation~\ref{eqn:artemova_potential}) 
is added to the $\ell=0$ moment. 

The BH is assumed to be inside the inner radial boundary of the computational domain.
The mass $M_{\rm bh}$ and angular momentum $J_{\rm bh}$ of this point mass are updated
at every time step with the material accreted through the inner radial boundary at $r=r_{\rm in}$
\begin{eqnarray}
\dot{M}_{\rm bh} & = & 2\pi r_{\rm in}^2\int d\Omega \left[\rho \max(0,-v_r)\right]\big|_{r_{\rm in}}\\
\dot{J}_{\rm bh} & = & 2\pi r_{\rm in}^2\int d\Omega \left[\rho j \max(0,-v_r)\right]\big|_{r_{\rm in}},
\end{eqnarray}
where the fluxes employed are those computed by the Riemann solver, which maximizes
conservative properties (e.g., \cite{F18}).
The initial values of $M_{\rm bh}$ and $J_{\rm bh}$ are obtained from the last \texttt{GR1D}
profile and set to the baryonic mass and 
angular momentum enclosed by the radius of the inner radial boundary 
at the beginning of the \texttt{FLASH} evolution. 
For simplicity, we do not consider the difference between
baryonic and gravitational masses. The instantaneous dimensionless BH spin 
parameter $a_{\rm bh}$ is then obtained from equation~(\ref{eqn:bh_spin}).
The updated values of $M_{\rm bh}$ and $a_{\rm bh}$ are then used to update the
gravitational potential of the BH (Equation~\ref{eqn:artemova_potential}).

The domain uses reflecting boundaries on the upper and lower $\theta$ edges, and
outflow boundaries at the inner and outer radial edges.
We use a logarithmicaly-spaced radial grid, and a polar grid equally spaced in
$\cos\theta$, as in \cite{FMM19}. The inner radial boundary is set so it falls
between the black hole event horizon and the ISCO radius. The domain extends across polar angles from 0 to $\pi$ with 112
cells, and the radial domain extends to $\sim 2\times$ the progenitor radius, 
depending on the progenitor, with 800 cells in total.   

As the BH accretes matter, its event horizon grows. When the
inner radial boundary falls below 130\% the event horizon radius, 
we excise an integer number of cells in the direction of increasing radius from the inner boundary,
setting the new inner radial boundary to be $r_{\rm in} \simeq r_{\rm h} + 0.75(r_{\rm isco}-r_{\rm h})$. 
Material in the excised cells is
assumed to be instantaneously accreted onto the BH, increasing its
mass and total angular momentum. An equal number of radial cells are added on
the outside of the domain and are filled with atmospheric material in order to keep
the total number of radial grid cells constant. The initial density, pressure and internal energy 
of this ambient material outside the star decreases as a power-law in radius.

In the case of the
\texttt{16TI} models, the accretion onto the black hole is small enough over the
timescale of the simulation that the inner boundary remains between the horizon
and ISCO for the duration of the simulation. In the case of the
\texttt{35OC\_SFHo} model, the accretion rate is significant over this timescale,
requiring the movement of the radial boundaries at multiple times throughout the
simulation.

We use a floor of internal energy, pressure, and density with radial and 
angular dependence, as described in Appendix \ref{sec:variable_floors}. 
Whenever the density floor is applied, the increase in matter is marked as atmospheric,
and has an electron fraction consistent with material in the cell before the floor
is applied. If the cell is in NSE, this is achieved by adding neutrons and
protons consistent with the desired electron fraction. If the cell is not in
NSE, $^{56}$Ni and either neutrons or protons are added consistent with the
desired electron fraction.

\begin{figure*}
\includegraphics*[width=\textwidth]{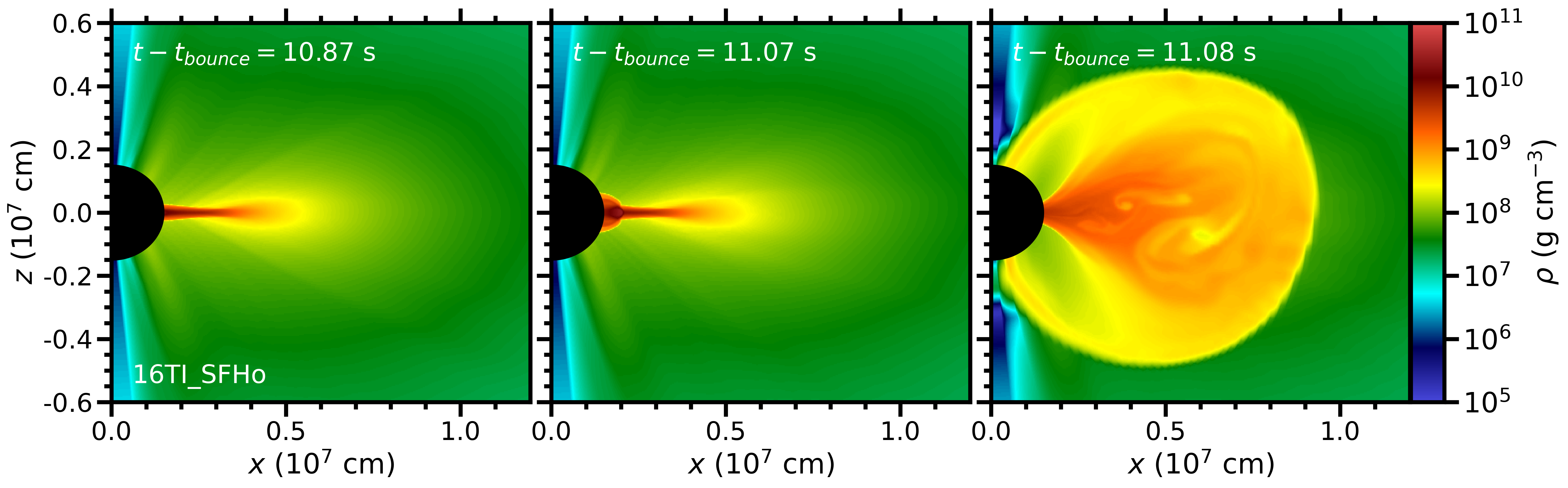}
\caption{Transition from \emph{dwarf disk} to thermalized disk in model \texttt{16TI\_SFHo}, with time after core bounce in the preceding {\tt GR1D} evolution labeled at the top of each panel. \emph{Left:} The shocked interface (dwarf disk) between supersonic inflows along the equator near the central BH. \emph{Center:} Pileup
of material and onset of the shock near the BH. \emph{Right:} Thermalized disk surrounded by a shock. The black circle at the origin is the excised inner radial boundary of the domain.}
\label{fig:dwarf_disk}
\end{figure*}

\subsection{Models evolved \label{sec:models}}

Table~\ref{tab:progenitors} shows all of the models evolved and the
key parameters being varied. Our baseline model \texttt{16TI\_SFHo} is the \texttt{16TI} progenitor evolved with the SFHo EOS in \texttt{GR1D}, and
thereafter evolved with \texttt{FLASH} using a viscosity parameter $\alpha=0.03$.

The dependence on the viscosity parameter is explored with models \texttt{16TI\_SFHo\_$\alpha$01} and \texttt{16TI\_SFHo\_$\alpha$001}, which use two additional values, $\alpha=\{0.1,0.01\}$, respectively. 
The sensitivity of the outflow to the density structure at BH formation is studied
with a model that uses the DD2 EOS in {\tt GR1D} (\texttt{16TI\_DD2}).
Finally, we evolve the \texttt{35OC} progenitor with otherwise default parameters (model \texttt{35OC\_SFHo}). 

Simulations are evolved until a time $t_{\rm max}$ after shock breakout from the surface of the
star, and until the shock front pressure exceeds $150$ dyn cm$^{-2}$ at the edge of the domain. 
This timescale varies for each model, and is shown in in Table \ref{tab:progenitors}.

\subsection{Outflow and shock analysis \label{sec:outflow} \label{sec:shock_fitting}}

Outflowing material is tallied by adding up 
unbound material over the computational domain at various times in the simulation. 
We use a positive Bernoulli parameter as a criterion to determine unbound status of a fluid element:
\begin{equation}
\label{eqn:bernoulli}
Be = \frac{1}{2} |\vect{v}|^2 + \epsilon + \frac{P}{\rho} + \Phi > 0,
\end{equation}
where $\vect{v} = \mathbf{v}_p + v_{\phi}\hat{\phi}$ is the full three dimensional velocity.
For reference, we assess the effect of using other unbinding criteria in Section \ref{sec:outflow_properties}.

We track the geometry of the shock that
bounds the accretion disk as it evolves. 
Initially, the shock front is detected by looking for a relative jump 
in pressure in the interior of the star, which we quantify
with a dimensionless pressure gradient parameter,
\begin{equation}
    H_p = \frac{r}{P} \pder{P}{r}.
\end{equation}
Further out radially, we use a velocity gradient parameter,
\begin{equation}
    H_{|\mathbf{v}_{\rm p}|} = \frac{r}{|\mathbf{v}_{\rm p}|} \pder{|\mathbf{v}_{\rm p}|}{r}.
\end{equation}
Finally, towards the surface of the star, and in models where the post shock material is well mixed, we use a threshold in $^{56}$Ni mass fraction ($X_{^{56}\rm{Ni}} > 10^{-8}$). Otherwise, we use the dimensionless pressure gradient parameter. Each shock detection begins from a prescribed radius, searching radially inward and recording the first instance in which the appropriate criterion exceeds a prescribed threshold value. 

We quantify the geometry of the shock front with a Legendre expansion (e.g., \cite{fernandez_2015})
\begin{equation}
\label{eq:legendre}
    r_{\rm s}(\cos\theta, t) = \sum_\ell a_\ell(t) P_\ell(\cos\theta),
\end{equation}
where $r_{\rm s}$ is the shock front radius at a given time for a given polar angle, $P_\ell$ are the Legendre polynomials, and $a_\ell(t)$ are the Legendre coefficients. 
We only consider the first three moments $\ell=\{0,1,2\}$, as they are the most informative regarding the evolution of the shock wave, with $a_0$ corresponding to the average shock radius, $a_1$ 
(dipole) describing the movement of the shock wave along the angular momentum ($z$-) axis, and $a_2$ (quadrupole) quantifying the relative extension in the polar versus equatorial direction. 

\section{Results \label{sec:results}}

\subsection{Overview of disk formation and evolution \label{s:disk_evolution}}

Following BH formation, the stellar material accretes 
radially at supersonic speeds, with an increasing asymmetry between polar and equatorial
regions due to centrifugal effects and the imposed angular dependence of $j$ 
in the progenitor (Section \ref{sec:flash}). 
As the BH mass increases, the circularization radius approaches the point
at which it crosses the ISCO radius (Figure~\ref{fig:radr_mgrv}), and
a high density region forms along the equator of the star, perpendicular to the angular momentum
vector (Figure \ref{fig:dwarf_disk}, left panel). Supersonic inflows of material from above and below the equator collide and create a shocked interface called
a \emph{dwarf disk} \cite{beloborodov_2001,lee_2006}, 
through which the flow still accretes supersonically into the BH. 
In some cases, this structure persists beyond the point at which the circularization radius
exceeds the ISCO radius (when the nuclear binding energy contributions to the
EOS are included) as shown in Figure~\ref{fig:radr_mgrv}.
Test simulations that ignore the nuclear binding energy and other source terms (i.e., adiabatic flow), skip an extended dwarf disk stage
and form the shocked disk at the expected point.

\begin{figure}
\includegraphics[width=\columnwidth]{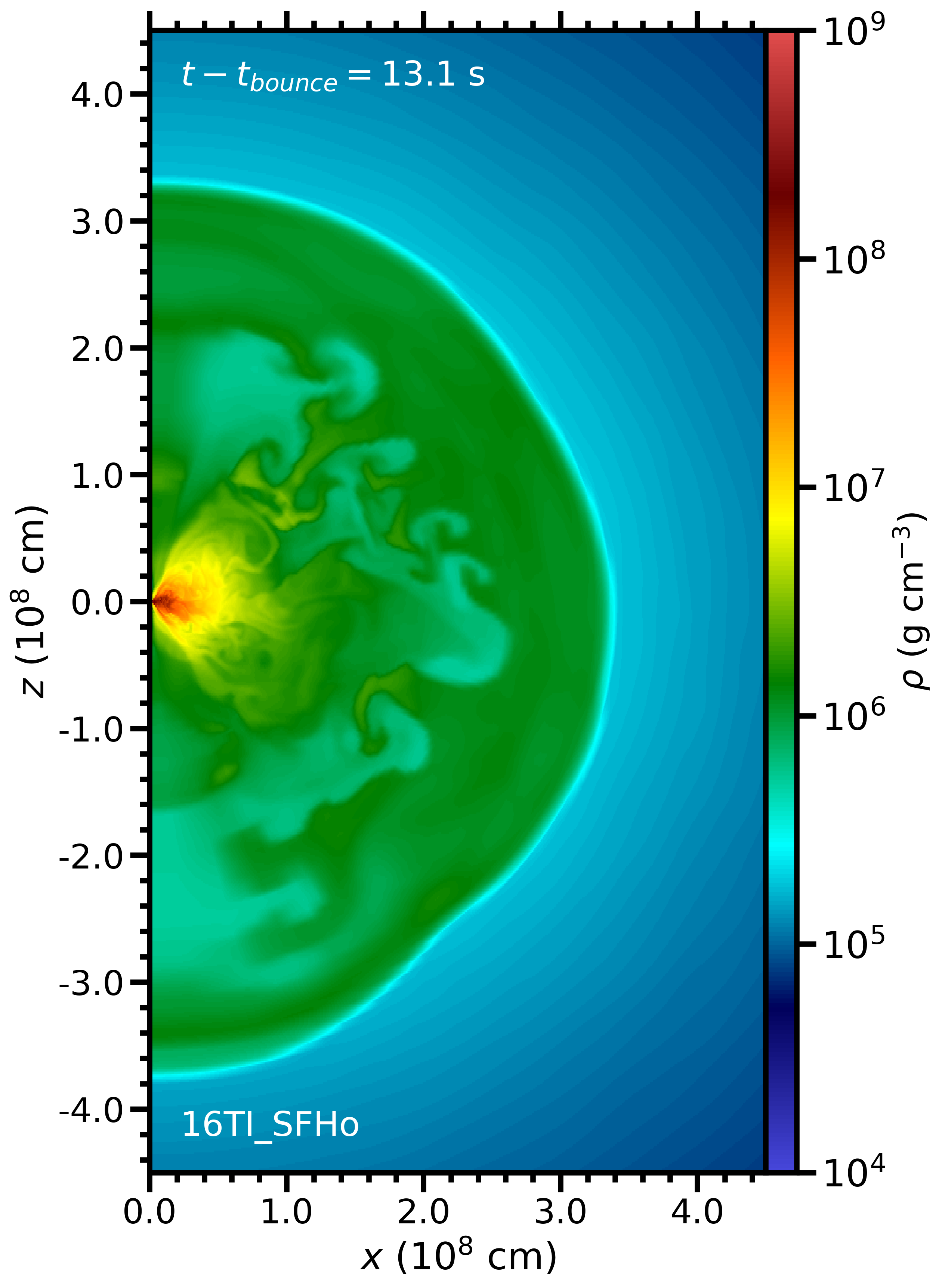}
\caption{Snapshot of the density distribution for model \texttt{16TI\_SFHo} at $\sim13.1$\,s post bounce. 
At this point in the simulation the shocked disk has formed ($11.1\,$s post bounce, Figure ~\ref{fig:dwarf_disk}), 
and a dominance of viscous heating drives turbulence and a disk wind which, combined with the
increasing specific angular momentum of accreted material, propels the shock out through the star. A slight north-south asymmetry and large-scale corrugation due
to oscillations (Figure~\ref{fig:legendre}) are already visible. These deviations from sphericity become more apparent as oscillations freeze out and the shock propagates through the star.}
\label{fig:density_colourplot}
\end{figure}

Eventually, material piles up in the equatorial region at a sufficient rate
to drive a shock out, inside which a thermalized accretion disk emerges on a timescale of $\sim 10$\,ms (Figure~\ref{fig:dwarf_disk}, see also \cite{mizuno_2004a, sekiguchi_2007,ott_2011, batta_2016}). 
The time at which this shocked bubble forms is marked in Figure~\ref{fig:radr_mgrv} by a vertical line. Two low-density funnels remain initially along the rotation axis, as material with insufficient angular momentum plunges directly into the BH. Eventually, the shock expands to cover all latitudes, 
as shown in Figure~\ref{fig:density_colourplot}.

\begin{figure}
\centering
\includegraphics[width=0.9\columnwidth]{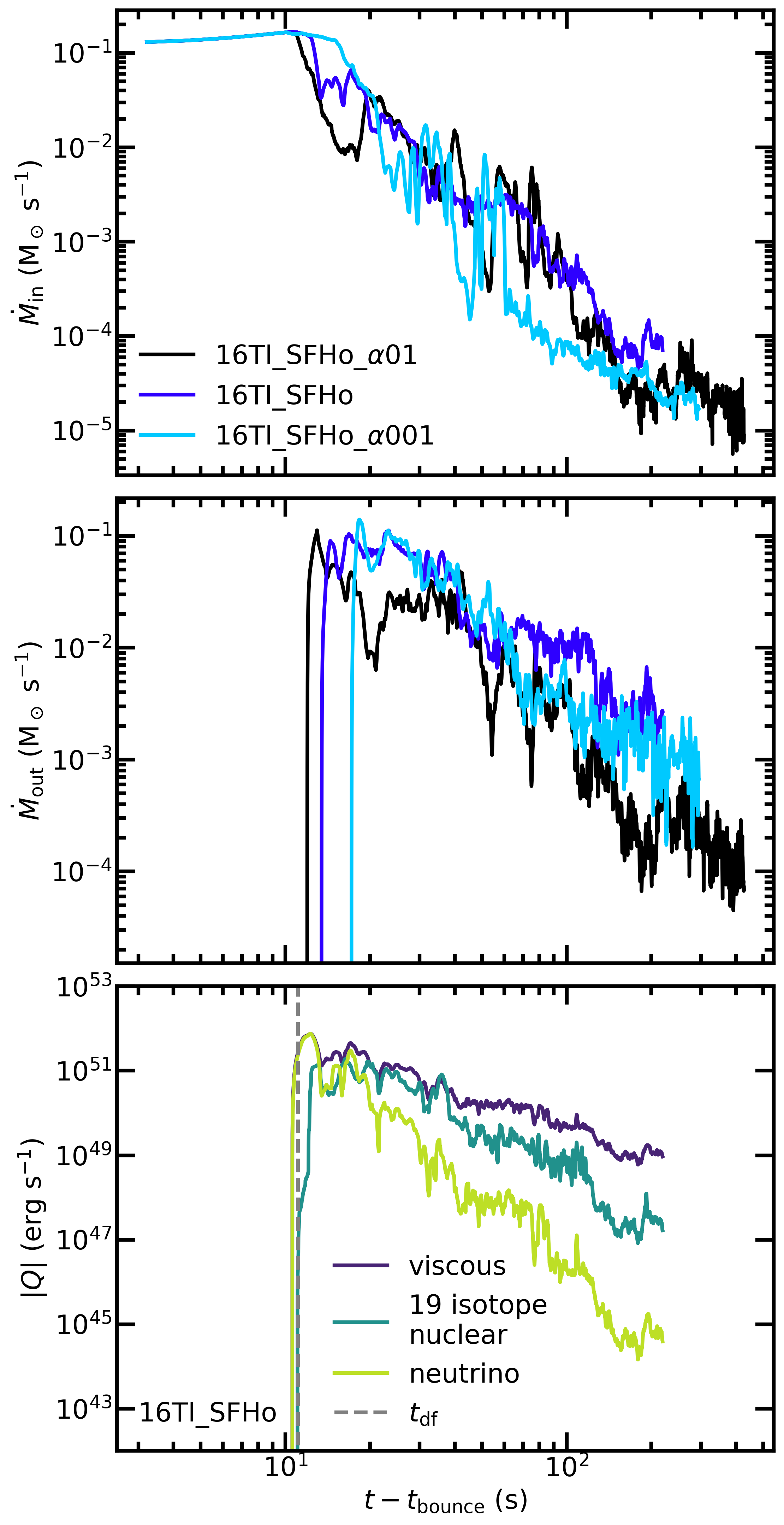}
\caption{\emph{Top:} Mass accretion rate across the inner radial boundary as a function of post-bounce time 
for selected models, as labelled. \emph{Middle:} Mass outflow rate with positive Bernoulli parameter across an extraction radius $r_{\rm  ej} = 10^{9}\,$cm, for the same set of models as the top panel. \emph{Bottom:} Total viscous heating, nuclear heating, 
and net neutrino cooling 
rates inside the shock radius in model \texttt{16TI\_SFHo}, as labeled. 
The \emph{19 isotope nuclear} curve shows the heating rate from the nuclear reaction network. 
Each curve has been smoothed with a moving average of width $0.5\,$s. The vertical dashed line shows the
time of shocked disk formation.}
\label{fig:heating_rates}
\end{figure}

\begin{figure}
\centering
\includegraphics[width=0.9\columnwidth]{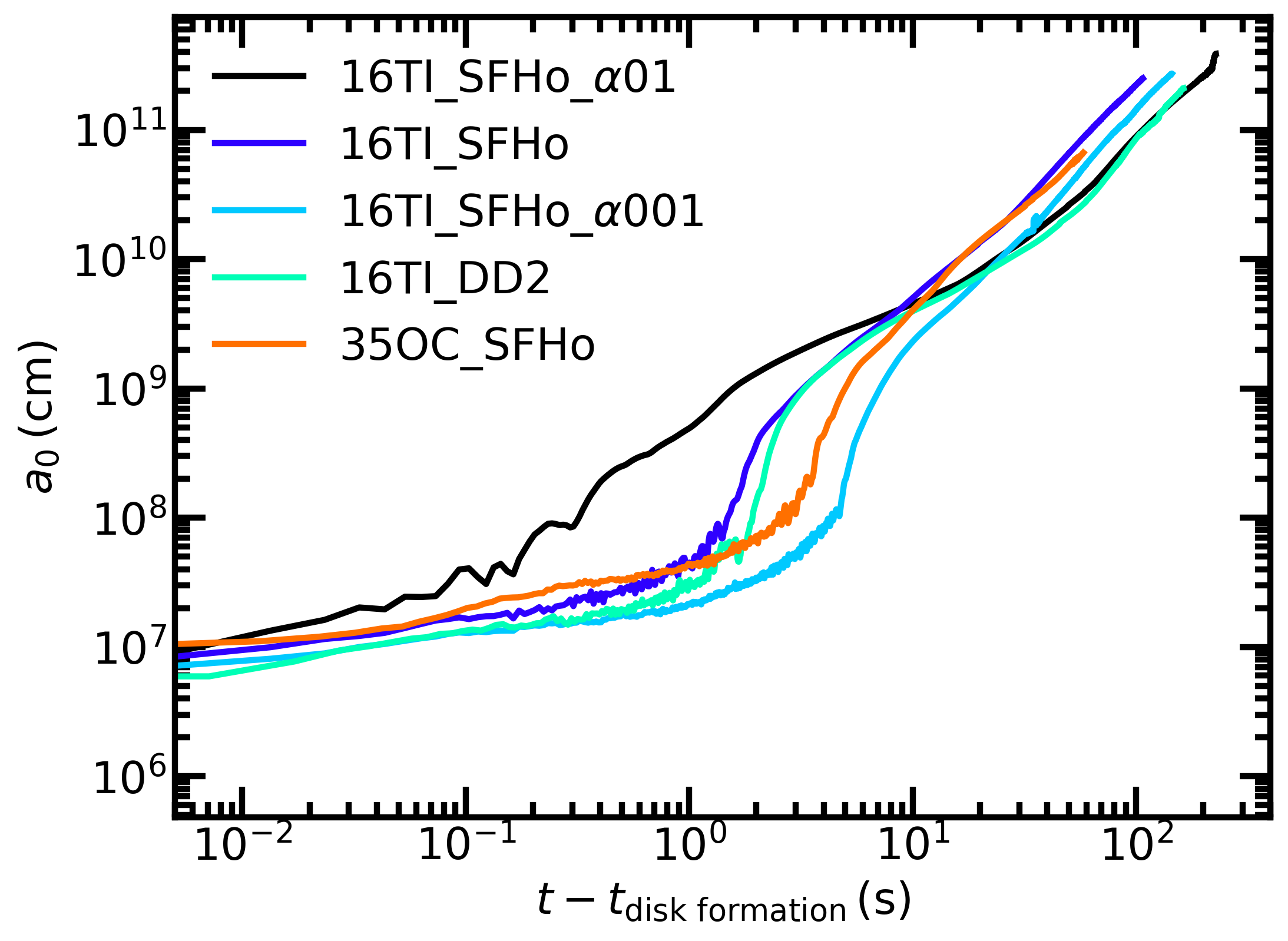}
\caption{Evolution of the average shock radius ($a_0$) as a function of time after disk formation (Table \ref{tab:progenitors}), for all models, until shock breakout from the stellar surface.}
\label{fig:average_shock_radius}
\end{figure}

\begin{figure}
\centering
\includegraphics[width=0.9\columnwidth]{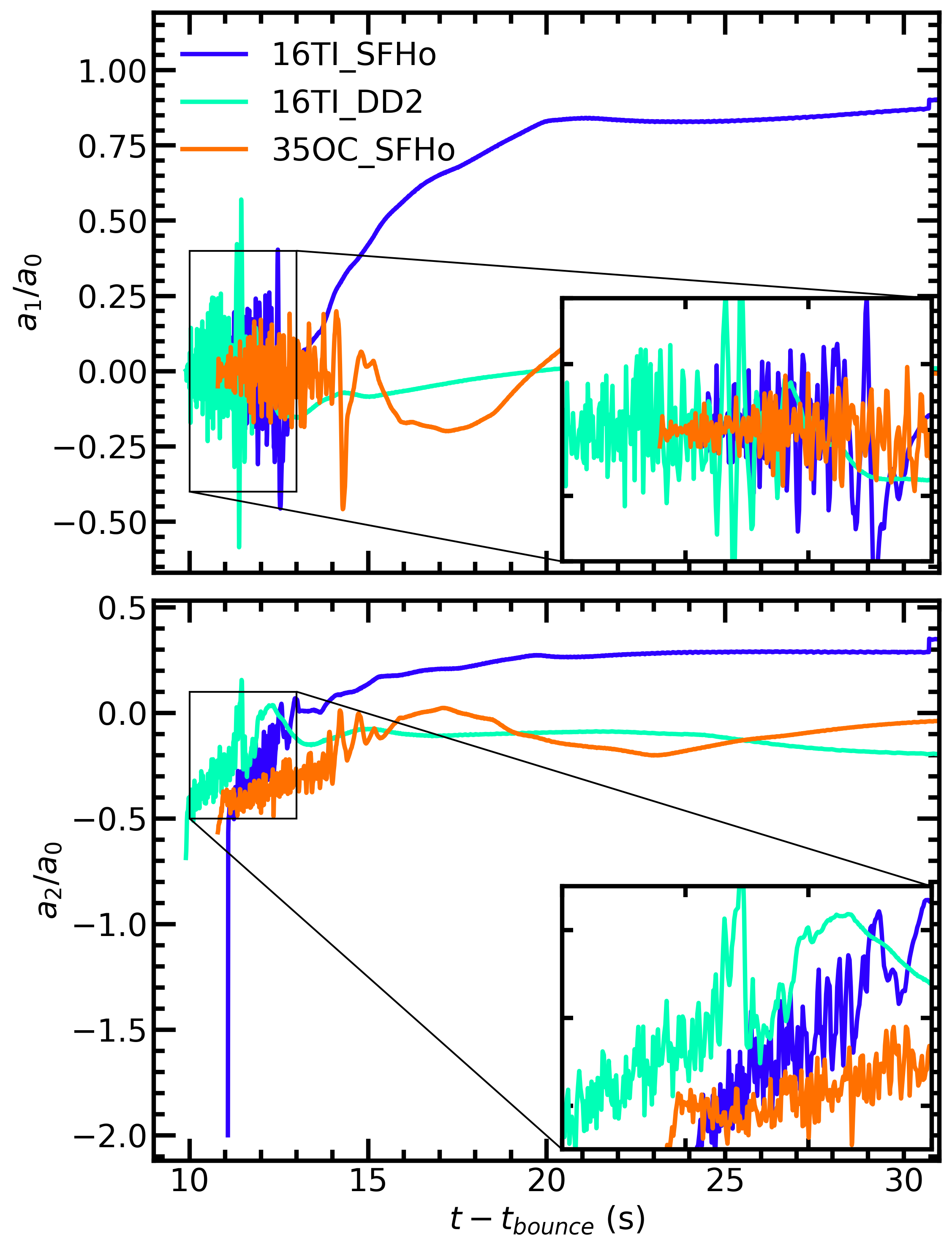}
\caption{Evolution of the normalized dipole (top) and quadrupole (bottom) Legendre coefficients of the shock radius
(Equation~\ref{eq:legendre}), for selected models. In most cases, the shock undergoes an initial oscillation phase before the geometry freezes, and continues to expand with little oscillation afterward. Each panel includes an inset which enlarges the time axis around the initial oscillation phase.}
\label{fig:legendre}
\end{figure}

Accretion to the BH decreases with time after shocked disk formation for the duration of the simulation, with
large stochastic fluctuations in some models, as shown in Figure~\ref{fig:heating_rates}.
Fluctuations are most clearly visible immediately after the formation of the shocked disk.
Accretion is mediated by viscous angular
momentum transport, with densities and temperatures high enough that neutrino
emission and absorption become dynamically relevant. Over a timescale of $\sim 1\,$s after disk formation in the \texttt{16TI\_SFHo} model, 
viscous and nuclear energy injection in the disk are approximately balanced by neutrino cooling (Figure \ref{fig:heating_rates}, bottom panel). This regime is referred to as Neutrino Dominated Accretion Flow (NDAF). As temperatures and densities drop in the disk as a result of the diminishing accretion rate, neutrino cooling drops off, and the energetics of the disk become dominated by viscous heating. This regime is referred to as Advection Dominated Accretion Flow (ADAF).
The interior of the shocked cavity becomes highly turbulent, 
as shown in Figure~\ref{fig:density_colourplot}. 

Figure~\ref{fig:average_shock_radius} shows the time evolution of the average shock radius after disk formation, for all models. Despite early oscillations, the size of the shocked disk increases 
monotonically with time over the duration of the simulation. 
The combination of net heating and increasing specific angular momentum 
of accreted material cause the shock to accelerate its expansion outward through the star 
in the ADAF phase. 
While the diminishing 
accretion rate with time from the collapsing star facilitates shock expansion as time elapses, the dominance of viscous heating over neutrino cooling is the main driver of this rapid expansion once the ADAF phase sets in.\footnote{
The shock that encloses the collapsar accretion disk is qualitatively different from that in slowly-rotating
core-collapse SNe, in which thermalization of accreting matter is offset by neutrino cooling and nuclear 
dissociation, leading to a stalled shock that responds sensitively to sudden changes in the accretion rate.
In core-collapse SNe, the cooling layer is supported by the protoneutron star and can thus remain at high densities for a long time, 
while in collapsars, significant disk cooling occurs only as long as the disk remains dense and hot enough.
}
The evolution of the average shock radius is non-monotonic with the strength of viscous angular momentum transport.
While the high-viscosity model \texttt{16TI\_SFHo\_$\alpha$01} initially expands more rapidly than the 
baseline model, it eventually slows down its expansion rate and ends up having the longest breakout time (Table~\ref{tab:progenitors}).

\begin{figure*}
\centering
\includegraphics[width=0.7\textwidth]{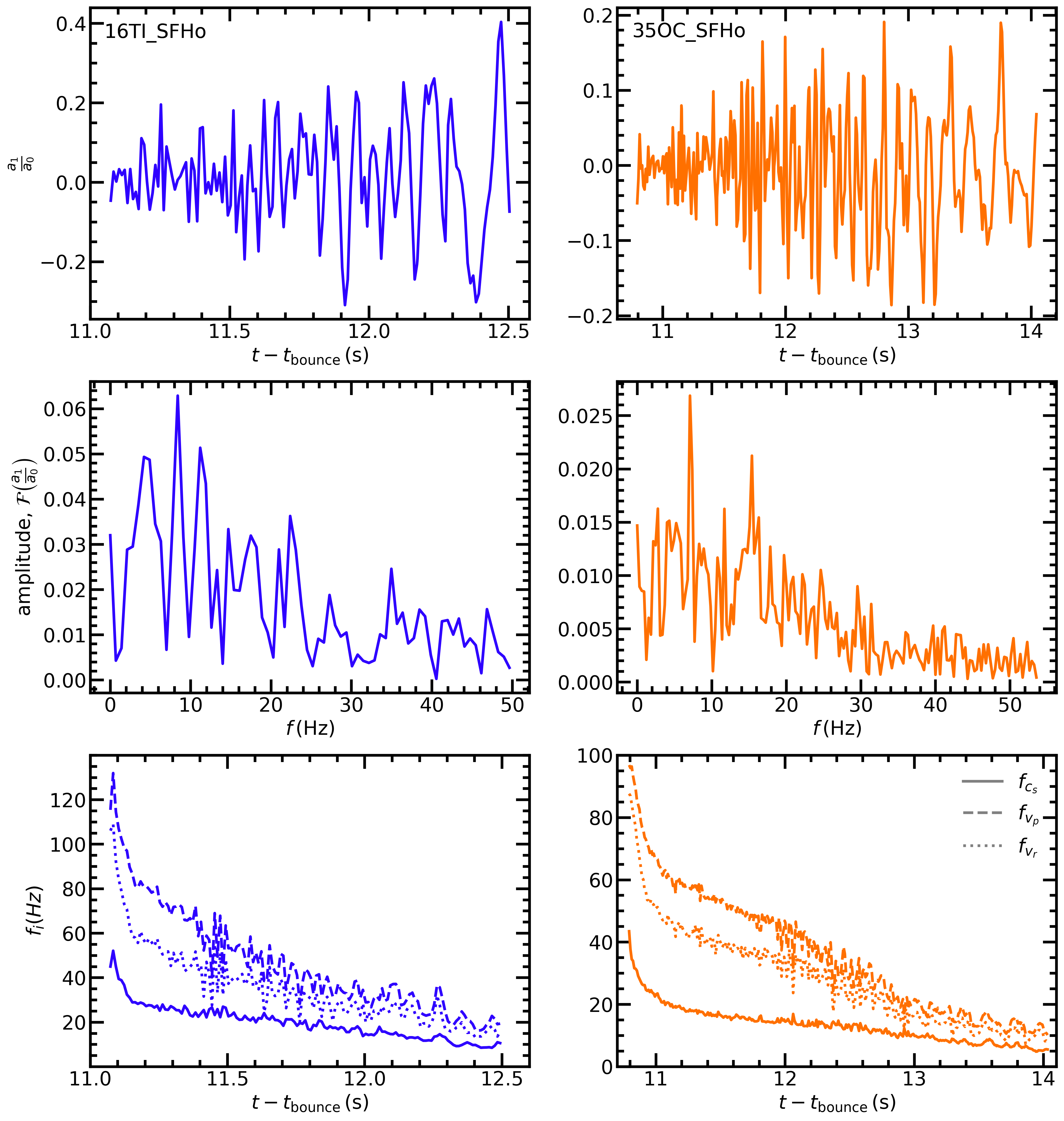}
\caption{Frequency analysis of initial shock oscillations. Data extends from the time of disk formation until the spherical harmonic coefficient begins to asymptote and the shock shape freezes. Left and right columns show data from different models, as labeled. \emph{Top:} Evolution of the Legendre coefficient $a_1$ normalized by the average shock radius $a_0$. \emph{Middle:} Amplitude (absolute value) of the Fourier transform $\mathcal{F}(a_1/a_0)$ of the normalized $\ell=1$ time series. \emph{Bottom:} Characteristic frequencies $f_i = v_i/a_0$ calculated using the average sound speed ($v_i=c_{\rm s}$), poloidal speed ($v_i=v_p$), and radial speed from behind the shock front ($v_i=v_{\rm r}$), as labeled.}
\label{fig:frequency_analysis}
\end{figure*}

Most models exhibit large scale shock oscillations over a timescale of several seconds following disk formation, 
after which the shock starts to rapidly expand. 
The oscillations are quantified in Figure~\ref{fig:legendre}, which shows the  time evolution (post-bounce) of the normalized
dipole ($a_1/a_0$) and quadrupole ($a_2/a_0$) moments of the shock surface (Equation~\ref{eq:legendre}).
Similar non-axisymmetric (spiral) shock oscillations were also reported by 
\cite{gottlieb_2022} in 3D GRMHD simulations without
neutrino cooling or nuclear energy changes. Accretion shocks around BHs are known to
be unstable to non-axisymmetric modes in both the isothermal and adiabatic limits \cite{molteni_1999,gu_foglizzo_2003,gu_lu_2006,
nagakura_yamada_2008,nagakura_yamada_2009}, although the stability properties with internal energy source
terms are less well studied than in the NS case (e.g., \cite{foglizzo_2007}).

In our models, the axisymmetric oscillations are concurrent with the NDAF phase. After the transition to the ADAF phase, rapid expansion starts, and oscillations stop. The high viscosity model \texttt{16TI\_SFHo\_$\alpha$01} skips the NDAF phase altogether, showing fewer early oscillations than the other models, with the shock expanding rapidly immediately after disk formation.

In models such as \texttt{16TI\_SFHo}, the shock bubble expands asymmetrically after a preferred polar direction is set once the disk oscillations freeze out (Figure \ref{fig:legendre}). In other cases,
such as the high-viscosity model \texttt{16TI\_SFHo\_$\alpha$01}, the shock expands roughly isotropically, with a slight predominance of the equatorial direction, reaching the surface with a slight extension to one pole. We surmise that the asymptotic
shock morphology arises as a combination of random oscillations frozen out as the disk becomes advective, the existence
and strength
of these oscillations given the balance of viscous heating versus neutrino cooling, and the
imposed angular dependence of the rotation profile in the star ($j\propto\sin^2\theta$), which results in an 
effective gravity that varies with angle and is weakest at the equator.

In Figure~\ref{fig:frequency_analysis} we show 
initial $\ell=1$ shock oscillations and its temporal Fourier spectrum for models 
\texttt{16TI\_SFHo} and \texttt{35OC\_SFHo} over $1-3$\,s after shock formation, along with
the characteristic frequencies $f_i = v_i/a_0$ associated with the shock crossing time
at various speeds $v_i$ (sound speed, average poloidal speed, and average radial speed). 
The Fourier amplitudes of model \texttt{16TI\_SFHo} show a broad peak
around $10$\,Hz, with power extending to $50$\,Hz. This range is consistent with that covered by the characteristic frequencies $f_i$, which decrease with time as the
shock cavity expands. A qualitatively similar result is obtained for
model \texttt{35OC\_SFHo}, but with shock oscillations occurring at
overall higher frequencies than in model \texttt{16TI\_SFHo}. We leave for future work 
a more thorough analysis of possible correlations between shock oscillations and temporal fluctuations
in the neutrino luminosity, as well as with gravitational wave emission.

Once oscillations freeze out, the shock expands through the remainder of the star with approximately constant shape. The post-bounce timescales for black hole formation, thermalized disk formation, and shock breakout from the stellar surface are listed in Table \ref{tab:progenitors}. 

\begin{table*}
\caption{Bulk outflow properties, obtained by integrating unbound material at the end of the simulation. Columns from left to right show model name, ejecta mass, ejecta kinetic energy at the end of the simulation, asymptotic ejecta kinetic energy (Equation \ref{eqn:Kinf}), mass-weighted average expansion velocity at infinity (Equations \ref{eqn:vinf}, \ref{eqn:ave_vinf}),
minimum electron fraction of outflowing material, $^{56}$Ni mass ejected, and SN light curve peak time (Equation \ref{eqn:tpeak}, \cite{arnett_1982}).
\label{tab:supernova}
} 
\begin{ruledtabular}
\begin{tabular}{lccccccc}
Model & $M_{\rm ej}$ (M$_\odot$) & $K_{\rm ej}$ ($10^{51}\,$ergs) & $K_\infty$ ($10^{51}\,$ergs) & $\langle v_\infty \rangle$ ($10^3\,$km/s) & 
  $Y_{\rm e,min}(t_{\rm max})$ & $M_{^{56}\rm{Ni}}$ (M$_\odot$) & $t_{\rm peak}$ (days) \\ 
\noalign{\smallskip} \hline
\texttt{16TI\_SFHo}              & 8.19 & 9.07 & 9.20 & 8.7 & 0.498 & 1.28  & 44.7  \\
\noalign{\smallskip}
\texttt{16TI\_SFHo\_$\alpha$01}  & 8.97 & 2.39 & 2.41 & 4.8 & 0.499 & 0.29 & 63.1 \\
\texttt{16TI\_SFHo\_$\alpha$001} & 7.93 & 4.34 & 4.37 & 6.0 & 0.481 & 0.81 & 52.9 \\
\texttt{16TI\_DD2}               & 9.17 & 3.67 & 3.70 & 5.6 & 0.500 & 0.63 & 59.2 \\
\texttt{35OC\_SFHo}              & 15.1 & 9.45 & 10.6 & 7.7 & 0.497 & 1.39 & 64.3 \\
\end{tabular}
\end{ruledtabular}
\end{table*}

\begin{figure*}
\includegraphics*[width=\textwidth]{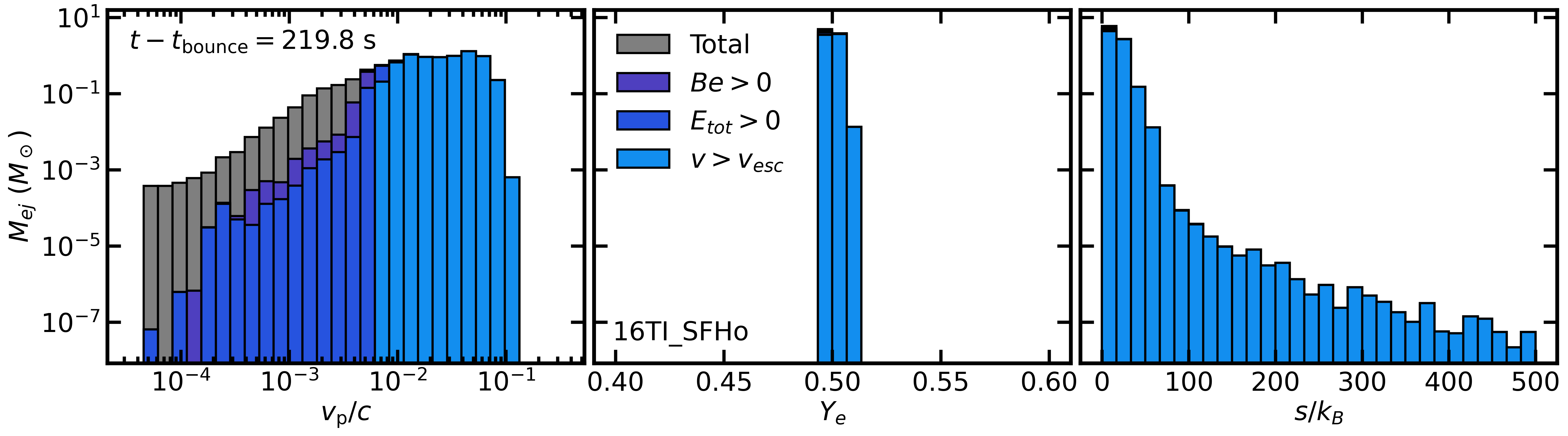}
\caption{Mass histograms of the outflow at the end of the simulation for model \texttt{16TI\_SFHo}, binned by poloidal velocity (left), electron fraction (center), and entropy per baryon (right). Different colours represent the gravitational binding criterion used, as labeled (\emph{Total} represents both bound and unbound matter).
}
\label{fig:criterion_histogram}
\end{figure*}

\begin{figure*}
\includegraphics*[width=\textwidth]{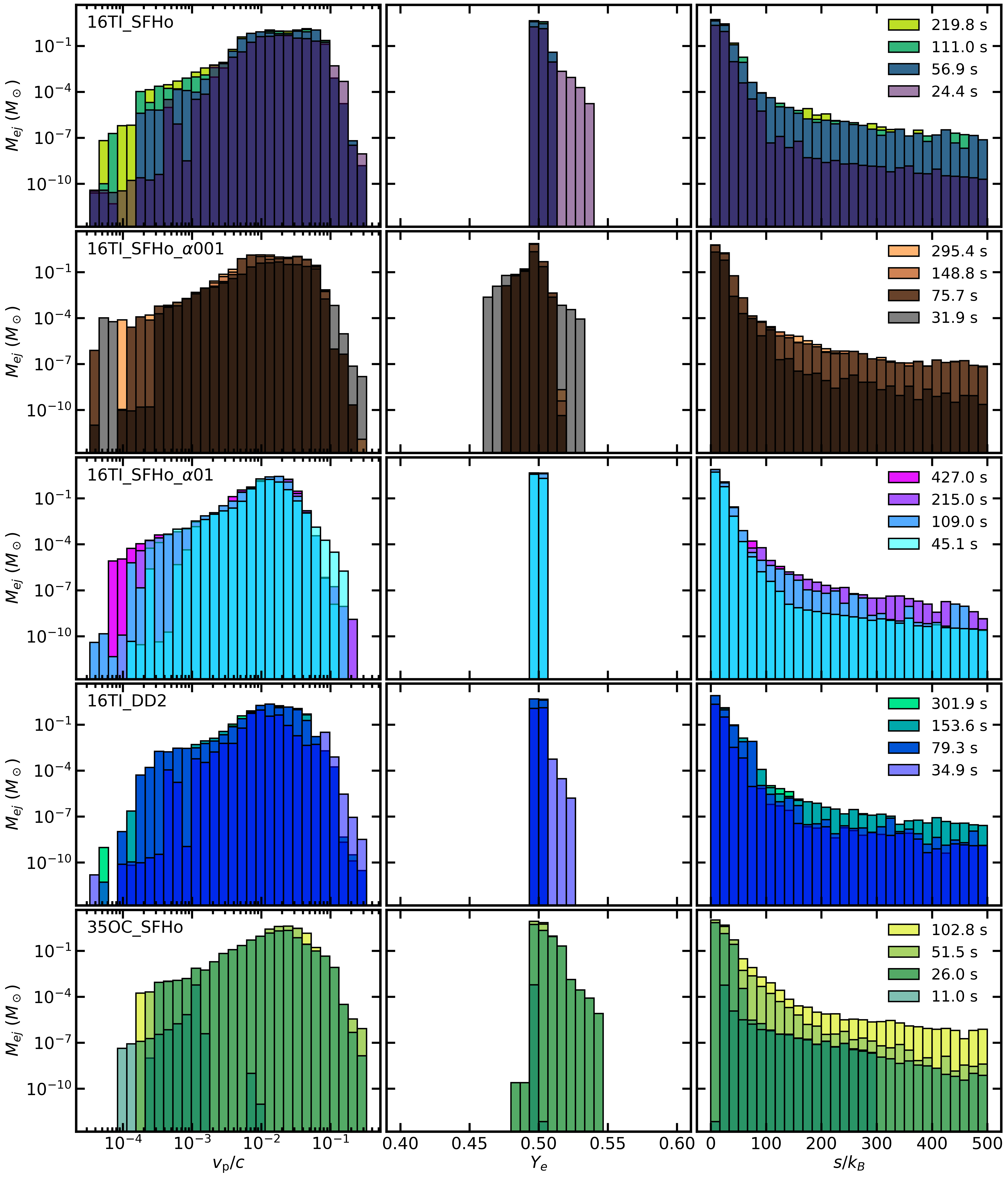}
\caption{Unbound mass histograms at various post-bounce times in each simulation, as labeled. Only material with positive Bernoulli parameter (Equation \ref{eqn:bernoulli}) and $v_r > 0$ is considered. Columns from left to right show histograms binned by poloidal velocity (left), electron fraction (center), and entropy per baryon (right).} 
\label{fig:all_model_histogram}
\end{figure*}

\begin{figure*}
\includegraphics*[width=\textwidth]{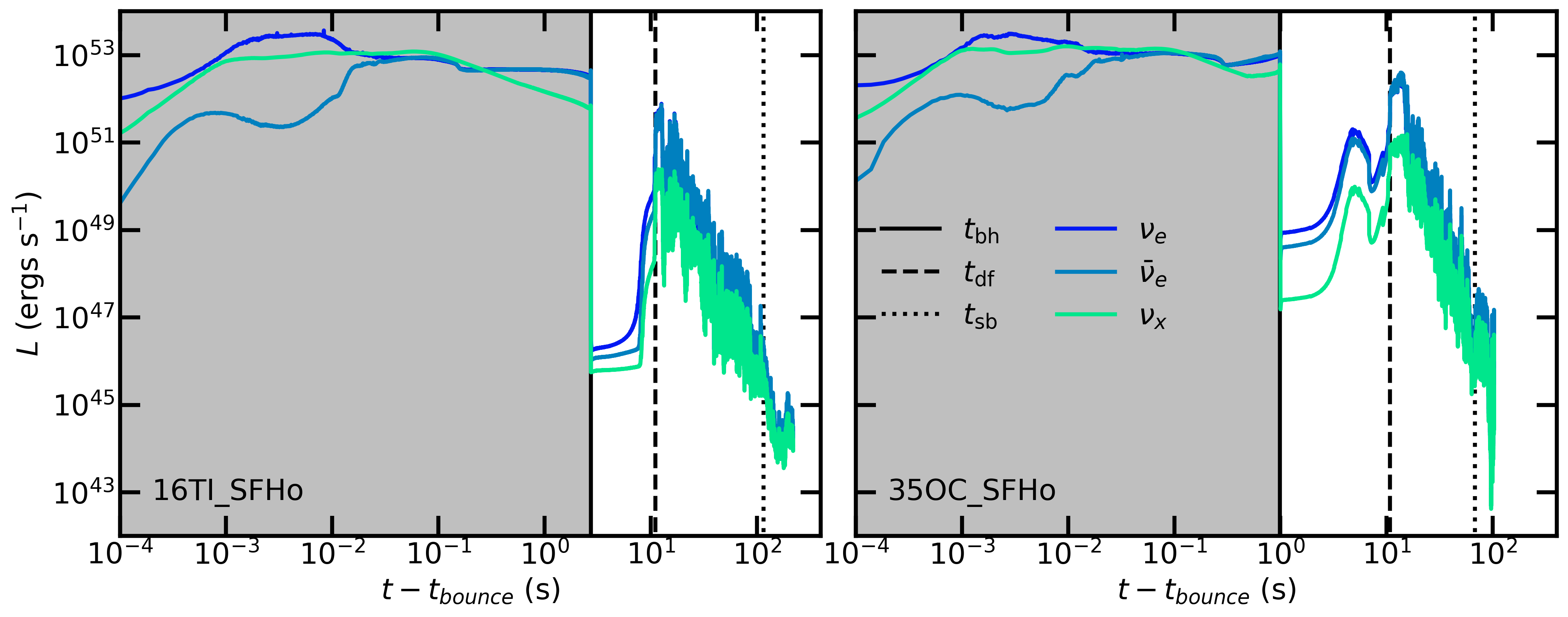}
\caption{Neutrino luminosities of $\nu_e$ [blue], $\bar{\nu_e}$ [teal], and $\nu_x$ [all heavy lepton species, green] as a function of post bounce time for selected models, as labeled. The vertical lines indicate the time of BH formation $t_{\rm bh}$ (solid black), shocked disk formation $t_{\rm df}$ (dashed black), and shock breakout from the star $t_{\rm sb}$ (dotted black). The shaded grey and white regions correspond to the \texttt{GR1D} and {\tt FLASH} portion of the evolution,
respectively.}
\label{fig:neutrino_luminosity}
\end{figure*}

\subsection{Outflow properties \label{sec:outflow_properties}}

Bulk properties of the disk outflow, obtained by integrating over unbound material at the end of the simulation,
are shown in Table~\ref{tab:supernova} for all models. The total ejecta mass has a monotonic
dependence on the strength of viscous angular momentum transport, with stronger viscosity leading to
more ejected mass. The ejecta kinetic energy at the end of the simulation $K_{\rm ej}$, 
on the other hand, shows non-monotonic behavior,
with a maximum for the baseline model \texttt{16TI\_SFHo} and the lowest value for the high-viscosity
case. Using the DD2 EOS before BH formation results in a slightly higher ejecta mass than the baseline model,
but with kinetic energy lower by a factor $\sim 2.5$. Changing the progenitor model to \texttt{35OC} results in 
a similar kinetic energy but almost double the ejecta mass than the baseline \texttt{16TI} model. 

Figure \ref{fig:criterion_histogram} shows the poloidal velocity, electron fraction, and entropy distributions of unbound material (section \ref{sec:outflow}) at the end of the simulation for model \texttt{16TI\_SFHo}, using different binding criteria: total speed exceeding the escape speed $v_{\rm esc}$, positive total specific energy $E_{\rm tot}$, positive Bernoulli parameter (Eq.~\ref{eqn:bernoulli}), and total ejecta (bound and unbound). The fastest ejecta ($v_{\rm p}\gtrsim 0.006\,$c) is unbound by kinetic energy alone, 
with a total mass of $7.2\,\Msun$. The remaining, slower ejecta has a significant internal energy component that contributes to its unbinding and which can be transformed into kinetic energy upon further expansion. 
The vast majority ($\sim 92\,$\%) of material ejected by the \texttt{16TI\_SFHo} model is unbound at the end of the simulation, according to the Bernoulli criterion. 
The ejecta velocity has a sharp cutoff at $\sim 0.2\,$c, consistent with BH accretion disks evolved in viscous hydrodynamics around NS merger remnants (e.g., \cite{FJ23}). The low velocity tail extends to $\sim 5\E{-5}\,$c.

The entropy distribution decays with increasing entropy, with a tail reaching several hundred $k_B$ per baryon. The fastest material that satisfies the escape velocity criterion dominates the distribution above $20-30\, k_{\rm B}$, with slower ejecta contributing mostly to the lowest entropy bin. Similar entropy distributions are obtained in viscous hydrodynamic simulations of BH accretion disks formed
in NS mergers, which produce most of their ejecta in the ADAF phase, driven by viscous heating and nuclear recombination (e.g., \cite{FJ23}).

The electron fraction distribution of our collapsar outflows is much narrower than that obtained in NS merger disk outflows. It has a peak at $Y_e\sim 0.5$, extending from $Y_e\sim0.49$ on the low end, up to $Y_e\sim0.51$ on the upper edge by the end of the simulation, with deviations from this general shape reflecting the degree to which neutrino interactions can neutronize disk material and possibly drive the $r$-process. There is no significant difference in $Y_e$ between ejecta components
with a different degree of gravitational binding, with more bound material contributing primarily with $Y_e\sim0.49$.

The mass outflow rate at $r=10^9$\,cm for selected models is shown in Figure~\ref{fig:heating_rates}. Curves follow a similar power-law decay structure with qualitatively similar peaks and dips as the mass accretion rate, shifted in time due to the interval needed for the ejecta to reach $r=10^9$\,cm.

Unbound mass histograms are shown in 
Figure~\ref{fig:all_model_histogram} for all models, considering only matter with positive Bernoulli 
parameter as well as $v_r>0$, and at several post-bounce times in the simulation.
Histograms have the same overall morphology as in Figure~\ref{fig:criterion_histogram}.

The electron fraction histograms are narrow in all cases, with variations between models limited to the interval $0.45 \lesssim Y_e \lesssim 0.55$. In models \texttt{16TI\_SFHo}, 
\texttt{16TI\_SFHo\_$\alpha$001}, and \texttt{16TI\_DD2}, the electron fraction distribution becomes narrower with time as neutrino luminosities drop off. The minimum electron 
fraction in the ejecta at the end of the simulation for each model $Y_{\rm e,min}$ is shown in Table~\ref{tab:supernova}. While there is a monotonic increase in minimum $Y_e$ 
with increasing viscosity, the variation in this quantity is less than $4\%$ for the viscosities used.

In each model, late-time ejecta contributes significantly to the high entropy tail of the distribution. During this time, the viscous heating rate drops 2 orders of magnitude while the peak density drops 5 orders of magnitude, leading to higher entropy ejecta at late times.

The velocity distribution is broadly similar between models, with some variation
at the low-velocity end. 

\subsection{Dependence of Disk Evolution on EOS and Progenitor Model}

The BH formation time relative to the bounce time in model \texttt{16TI\_DD2} is $\sim 2.5\,$s longer than in model \texttt{16TI\_SFHo}, as expected for a stiffer EOS. 
The longer evolution time implies that the outer stellar layers at the same Lagrangian mass coordinate 
have collapsed to a deeper radius in model \texttt{16TI\_DD2} than in \texttt{16TI\_SFHo}, 
with the disk forming at an earlier time post-bounce.
Both \texttt{16TI\_SFHo} and \texttt{16TI\_DD2} go through an NDAF phase and exhibit early shock oscillations, 
before viscous heating becomes dominant and shock expansion ensues. 
The average shock radius in model \texttt{16TI\_DD2} starts out smaller than in 
the baseline model (Figure \ref{fig:average_shock_radius}), but upon transition to the ADAF phase, the shock in  model \texttt{16TI\_DD2} accelerates to match
the position of that from \texttt{16TI\_SFHo}, eventually falling 
behind, having a lower kinetic energy and
longer breakout time (Table \ref{tab:progenitors}). 
 This difference in evolution can be traced back to the longer post bounce time to BH formation in model \texttt{16TI\_DD2}. 
The presupernova star has two prominent discontinuities in the angular momentum profile, corresponding to the lower and upper edges of the 
silicon burning shell outside the iron core. During the \texttt{GR1D} evolution, the angular momentum profile is stretched radially as the star collapses, 
with the outermost discontinuity becoming a broad dip in the angular momentum profile at $\sim 10^8-10^9\,$cm in the \texttt{16TI\_SFHo} model. Due to the longer collapse time of the \texttt{16TI\_DD2} model, the dip in angular momentum is flattened out. This results in a lower mass accretion rate at late times ($\sim 70-130\,$s) in model \texttt{16TI\_DD2} due to the higher angular momentum of material being added to the disk, reducing the energy injection by the disk wind, and ultimately delaying shock breakout from the surface of the star. 

In model \texttt{35OC\_SFHo} the progenitor star is much more massive at the end of its life (for \texttt{35OC}, a mass: $28.1\,\Msun$, for \texttt{16TI}: $13.9\,\Msun$) than the fiducial progenitor, while being smaller in size and thus much more compact (for \texttt{35OC}, a radius: $1.6\E{11}\,$cm, for \texttt{16TI}: $7.3\E{11}\,$cm). The evolution of model \texttt{35OC\_SFHo} is faster than the fiducial model, with BH formation, disk formation, and shock breakout occurring on shorter timescales. Due to the high accretion rate in model \texttt{35OC\_SFHo}, we need to move the inner radial bound several times before thermalized disk formation slows down the BH accretion rate. This model exhibits similar early oscillations of the shock before the onset of rapid expansion to the fiducial model. Notably, in predicting the Lagrangian mass coordinate for accretion disk formation, the circularization radius $r_{\rm circ,A}$ exceeds the ISCO of the BH at two points (Figure~\ref{fig:radr_mgrv}), with the disk formation point we obtain being consistent with the second crossing. This pattern appears in the neutrino luminosity (Figure \ref{fig:neutrino_luminosity}) as a bump before thermalization of the dwarf disk. Model \texttt{35OC\_SFHo} produces significantly more ejecta than the fiducial model, but with comparable kinetic energy.

\begin{figure*}
\centering
\includegraphics[width=0.8\textwidth]{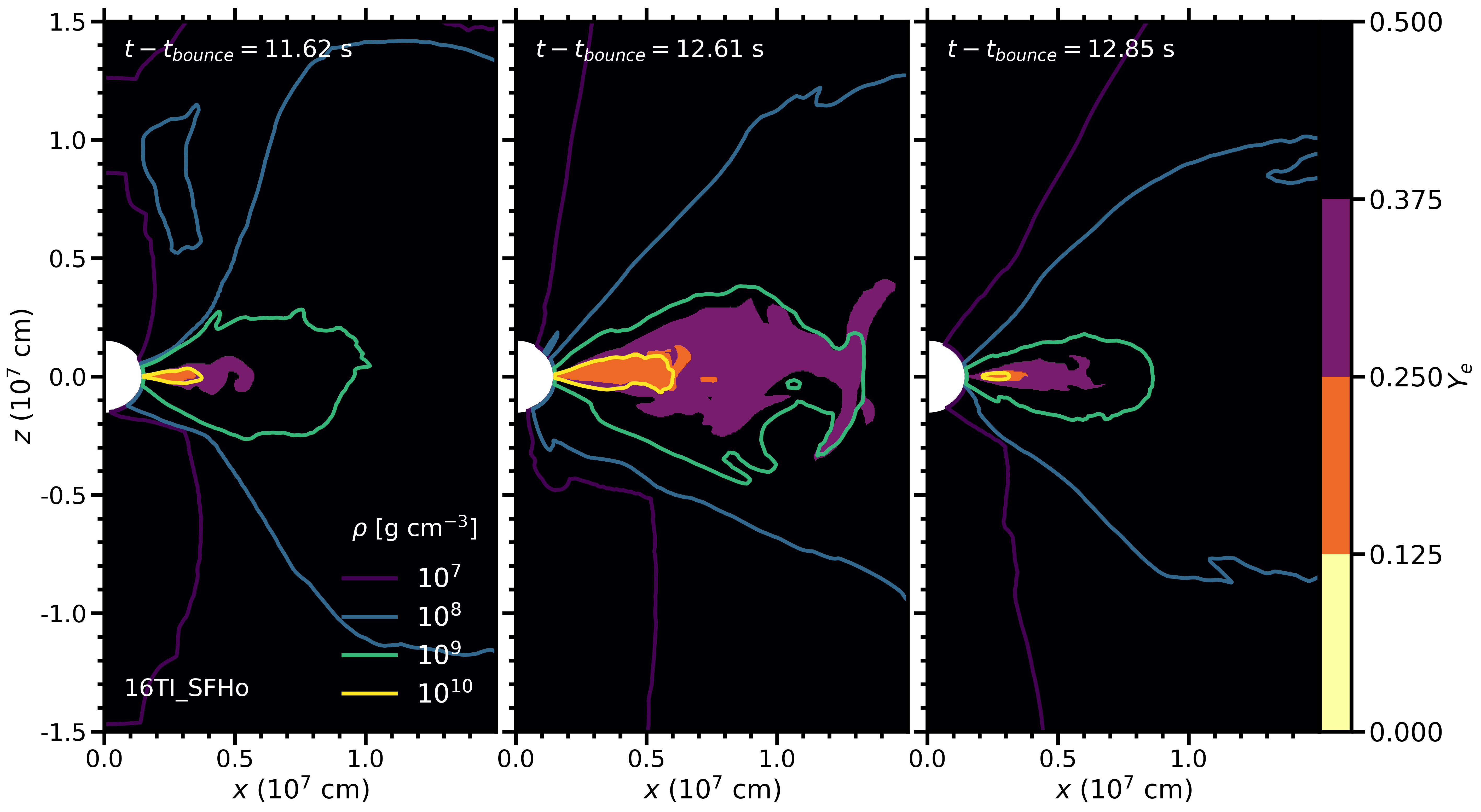}
\caption{Snapshots of the electron fraction within the inner disk in model \texttt{16TI\_SFHo} around the time of maximum neutrino emission (c.f. Fig.~\ref{fig:neutrino_luminosity}), with overlayed density contours, as labeled.}
\label{fig:electron_fraction}
\end{figure*}

\subsection{Neutrino emission and neutronization}

After the formation of the shocked disk, densities and temperatures are
high enough for charged-current weak interactions to become important in cooling
the disk and changing its composition. In particular, since material from the
collapsing star has $Y_e\simeq 0.5$, any path to $r$-process nucleosythesis requires
a significant amount of electron-type neutrino/antineutrino 
emission and absorption, in order to increase the ratio of neutrons to protons
toward its neutron-rich equilibrium value for a disk with partially-degenerate
electrons.

Figure \ref{fig:neutrino_luminosity} shows total neutrino luminosities of $\nu_e$, $\bar{\nu}_e$, 
and $\nu_x$ (which represents all heavy lepton species) for selected models. During the 
protoneutron star phase, luminosities rise steeply following shock breakout from the
neutrinosphere, thereafter decreasing more gradually (factor of $\sim 5$), followed by a sharp drop 
as the NS collapses to a BH. 

As material from the infalling star accretes onto the BH, densities and temperatures gradually increase 
toward the equatorial plane due to centrifugal effects, leading to a slow increase in luminosities. As the dwarf disk forms (Figure~\ref{fig:dwarf_disk}), luminosities of all flavors accelerate their rise. Formation of the thermalized disk marks a sharp increase in luminosities due to the higher densities and temperatures. 

This delay time between the sharp drop in neutrino luminosities at BH formation ($t_{\rm bh}$ in Figure~\ref{fig:neutrino_luminosity}) and the spike shortly after shocked disk formation ($t_{\rm df}$) depends on the angular momentum profile of the star, and can be a useful observational diagnostic of the rotational profile of collapsar progenitors. The luminosity maximum after thermalized disk formation depends on the thermodynamics of the disk when it forms, which in turn depends on the compactness of the star $M/R$ and on the accretion rate $\dot{M}_{\rm acc}$. After reaching a peak, luminosities decay as a power law in time. 
This decay is related to the difference between the accretion rate onto the BH (which depends on the angular momentum transport rate in the disk) and the rate at which mass crosses the shock and feeds the disk, which ultimately depends on the radial dependence of density and angular momentum of the star. 

Detecting these observational signatures in neutrinos would require a galactic collapsar, with current capabilities. 
This is unfortunately limited by the galactic collapsar rate of $\sim 1$ per $10^4$ years (e.g., \cite{graur_2017}). 

The lack of neuton-rich material in the outflow (Fig.~\ref{fig:all_model_histogram}) could
be seen to be at odds with the substantial neutrino emission produced by all models (Figure~\ref{fig:neutrino_luminosity}). The answer is provided in Figure~\ref{fig:electron_fraction}, which shows the electron fraction in the inner accretion
disk for our baseline model, around the time of peak neutrino emission. 
Over a timescale of $\sim 1\,$s around the maximum in neutrino emission, material in the densest regions of the accretion disk neutronizes to $Y_e<0.25$. The vast majority of this material is fully accreted onto the BH, however, and does not contribute to the outflow except possibly through trace amounts mixed into the shock cavity. As accretion continues and the density in the disk drops, neutrino emission decreases from its maximum and so does the degree of neutronization, with the electron fraction remaining closer to $Y_e\sim 0.5$ as the accretion disk is continually fed by infalling stellar material. By the time shock expansion accelerates in the ADAF phase, giving rise to an outflow, there is negligible neutronization of post-shock material.

\subsection{An engine for type Ic-BL supernovae?}

In order to explain type Ic-BL SNe with the collapsar disk outflow alone, not only does the explosion need to be successful -- shock breaking out of the stellar surface with enough energy -- but also sufficient $^{56}$Ni must be produced in order to power the light curve over a timescale of months \cite{macfadyen_2003}. 
The average $^{56}$Ni mass for a sample of type Ic-BL was found to be $0.32M_\odot$ by 
\cite{lyman_2016} through bolometric light curve fits, although values
can be as high as $0.7M_\odot$ (e.g., for SN 1998bw \cite{iwamoto_1998,woosley_1999}), and low-$^{56}$Ni mass events
could be missed due to selection effects \cite{ouchi_2021}.
An analysis of a large sample of type Ic-BL SN spectra by \cite{modjaz_2016} found mean line width velocities in the range $13,000-21,000$\,km\,s$^{-1}$ ($0.04-0.07$\,c), depending on whether the SN was accompanied by a GRB, and the epoch at which it was measured. 

To assess the plausibility of our collapsar disk outflows as engines of these SNe, we integrate unbound outflowing $^{56}$Ni (obtained from the 19-isotope network and NSE solver) at the end of the simulation. Table \ref{tab:supernova} lists nickel masses as well as total ejecta masses. All models produce sufficient $^{56}$Ni to power a generic type Ic-BL SN light curve, with variation in $^{56}$Ni yield between models spanning a factor of $\sim 4$. 
 
We obtain the asymptotic expansion velocity $v_\infty$ at infinity by equating the asymptotic kinetic energy per unit mass to the Bernoulli parameter, which implies full conversion of internal energy to kinetic energy through adiabatic expansion of unbound material, with no additional energy sources:
\begin{equation}
\label{eqn:vinf}
    \frac{1}{2}v_\infty ^2 = \max(Be,0),
\end{equation}
where $Be$ is given by Equation~(\ref{eqn:bernoulli}). We then compute an asymptotic ejecta
kinetic energy as
\begin{equation}
\label{eqn:Kinf}
    K_\infty = \int \frac{1}{2}v_\infty^2\, dM_{\rm ej}.
\end{equation}
The ejecta-mass-weighted average velocity of unbound material is defined as
\begin{equation}
\label{eqn:ave_vinf}
    \langle v_\infty \rangle = \frac{\int v_\infty dM_{\rm ej}}{\int dM_{\rm ej}},
\end{equation}
where, as implied by Equation~(\ref{eqn:vinf}), the integral  
is carried out over all cells that satisfy $Be>0$. 

Table \ref{tab:supernova} shows that the average outflow velocity in the ejecta from our simulations is systematically lower, by a factor of at least $\sim 2$, than what is inferred from the spectra of type Ic-BL supernovae. Note however that our asymptotic ejecta kinetic energies $K_\infty$ are consistent with the range inferred for this SN subclass \cite{lyman_2016}, thus the lower average
velocities can be a consequence of the larger ejecta masses we find (by a factor $2-3$) compared to the average value for type Ic-BL SNe.

For reference, we also estimate the peak time for a SN light curve using \cite{arnett_1982}
\begin{equation}
\label{eqn:tpeak}
    t_{\rm peak} = \left( \frac{3}{4\pi} \frac{\kappa M_{\rm ej}}{\langle v_\infty \rangle c}\right)^{1/2}.
\end{equation}
assuming an opacity of $\kappa = 0.1\,$cm$^2\,$g$^{-1}$. Table \ref{tab:supernova} shows the resulting rise times. The larger ejecta masses and lower asymptotic velocities drives the peak
time toward higher values that what would be obtained with average values for the Ic-BL class.

\subsection{Comparison to recent work}

While multi-dimensional global collapsar simulations have a long history 
\cite{bodenheimer_1983,macfadyen_1999,proga_2003a,mizuno_2004a,mizuno_2004b,fujimoto_2006,nagataki_2007,sekiguchi_2007,harikae_2009a,harikae_2010b,lopezcamara_2009,lopezcamara_2010,ott_2011,sekiguchi_2011,batta_2016,obergaulinger_2017,nagataki_2018,aloy_2021,gottlieb_2022,janiuk_2023,shibata_2023,crosato_2023}, only recently have
models been developed which simultaneously include (1) global star collapse with self-consistent 
disk formation and subsequent accretion and outflow, 
(2) angular momentum transport, and (3) neutrino emission and absorption with appropriate
microphysics and evolution of $Y_e$ \cite{just_2022,fujibayashi_2022,fujibayashi_2023}, thus we
focus our comparative discussion on these recent studies.


Ref. \cite{just_2022} use Newtonian hydrodynanics with a pseudo-Newtonian BH, the same components
of the viscous stress tensor with similar viscosity strengths, and the \texttt{16TI} progenitor. 
The two main qualitative differences with our models are their use of an energy dependent
M1 neutrino transport (whereas we use a gray leakage scheme with lightbulb absorption), 
and the initial condition for the simulations, which
are set up by placing a BH at the center of the star at the time of core bounce (whereas we evolve with {\tt GR1D} until
BH formation). In addition, they do not include the energy input from a nuclear reaction network.

Disk formation occurs $\sim 1-2\,$s earlier in our simulations than in the corresponding models of \cite{just_2022}, likely stemming from the difference in initial condition. Figure \ref{fig:radr_mgrv} predicts disk formation at Lagrangian enclosed mass of $\sim 3.2\Msun$ for the 16TI progenitor and the SFHo EOS, whereas \cite{just_2022} predicts BH formation at $\sim 3.8 \Msun$. 
Similarly our BH masses at disk formation are $\sim 0.3\Msun$ lower. We see the same dependence of $Y_{\rm e,min}$ in the ejecta with increasing viscosity, pushing the minimum electron fraction towards $Y_e = 0.5$. We also see the same monotonic decrease in final black hole masses with increasing viscosity, but with our values being $\sim 0.5-1\Msun$ lower. Like \cite{just_2022}, we see the monotonic relationship of the NDAF phase duration with viscosity, before advection and viscous heating become dominant during the ADAF phase (the intermediate viscosity value resulting in the shortest NDAF phase, the low viscosity model having the longest NDAF phase, and the high viscosity model immediately starting in the ADAF phase). We find (non-monotonic) explosion energies consistent with those of \cite{just_2022}. 

Notably, \cite{just_2022} also finds a non-monotonicity in the shock breakout time with viscosity: their intermediate viscosity run is the fastest, followed by the high-, and finally low viscosity, whereas we find that the intermediate viscosity is fastest, followed by low- and finally high viscosity. Unlike \cite{just_2022}, however, the geometry of our shock waves at the time of shock breakout in the low- and intermediate viscosity models are extended to one pole, while the high viscosity model is more spherical. Instead, \cite{just_2022} find that the low- and intermediate viscosity runs are nearly spherical, while the high viscosity model is equatorially extended. 

A comparison study of viscous hydrodynamic evolution of NS merger accretion disks \cite{fernandez_2023} has shown
that M1 transport results in more efficient cooling than the leakage scheme used in our {\tt FLASH} setup, which depends on the
adopted local prescription for the optical depth. This inefficient cooling is also evident when comparing our
scheme with time-independent Monte Carlo transport on simulation snapshots \cite{fahlman_2022}. While this inefficient
cooling can in principle affect neutronization of the disk, our overall agreement with the results of \cite{just_2022} shows
that for viscous hydrodynamic evolution, for which outflow occurs in the ADAF phase, neutrino transport differences are not 
consequential for the occurrence of the $r$-process in the outflow, and play a sub-dominant role in mass ejection.

Ref. \cite{fujibayashi_2022} evolves the collapse of several rotating helium and Wolf-Rayet progenitor stars in axisymmetric numerical relativity, using M1 neutrino radiation transport, and a turbulent length scale to parameterize the strength of viscosity. While their progenitors are not directly comparable to ours, 
they start from a pre-collapse progenitor, and follow the evolution to bounce and BH formation before forming the disk. 
They also find insufficient neutronization to support the production of $r$-process elements in all their models. The entropy 
distribution of the ejecta extends to several hundred $k_B$, like in our models. Despite the differing progenitor models, we find similar disk outflow energies, while our ejecta masses are larger by a factor of several. This is likely due to the shorter duration of their models, and the use of an extraction surface instead of integrating over the entire domain at the end of the simulation (thus not accounting for mass outside the extraction threshold that may become unbound after crossing it). As a result of these smaller ejecta masses, the estimated supernova light curve rise time is shorter than those we estimate here, by a factor of a few. 

Models from \cite{fujibayashi_2022} are broken down into two qualitative groups, according to their evolution. First, those that have a higher infall rate at the time of disk formation, which undergo a NDAF phase before viscous heating becomes dominant over neutrino cooling, and evolve in a qualitatively similar way to our low- and intermediate viscosity models. Models with lower infall rate at the disk formation time are such that viscous heating dominates over neutrino cooling over the entire disk expansion. This is qualitatively similar to our high viscosity model. 

Ref. \cite{fujibayashi_2023} follows the disk outflow in three different progenitor stars with high core compactness, varying viscosity, rotation rate, and resolution, among other parameters, and using the same method as Ref. \cite{fujibayashi_2022} but now placing a BH at the center of the star at the time of core bounce. Models are run for values of $\alpha = 0.03,0.06,0.10$. Like our high viscosity case, their high-$\alpha$ run proceeds with no NDAF phase, leading to the outflow starting a short time after disk formation. Their low viscosity run evolves in a qualitatively similar way to our low and intermediate viscosity models, where the explosion is initially delayed due to the presence of an NDAF phase. While the models they run are not directly comparable to ours, we find the same monotonic increase in ejecta mass with increasing viscosity. It is unclear if they see the same non-monotonicity in shock breakout time with viscosity, since their models are run for only $\lesssim 20\,$s of simulation time.


\section{Summary and Discussion \label{sec:summary}}

We have studied the long-term outflows from accretion disks formed in rotating Wolf-Rayet stars undergoing core collapse. We evolve the progenitor from core-collapse to BH formation in spherical symmetry using {\tt GR1D} (Figure~\ref{fig:radr_mgrv}), and thereafter in axisymmetry using {\tt FLASH}. A shocked, centrifugally-supported disk emerges self-consistently in our simulations (Figures~\ref{fig:dwarf_disk}-\ref{fig:density_colourplot}), and is subject to angular momentum transport via shear viscosity, and heating/cooling due to viscosity, neutrino emission and absorption, and nuclear energy release (Figure~\ref{fig:heating_rates}). Unbound mass is ejected from the disk once it enters an ADAF stage with sub-dominant neutrino cooling. Our main results are the following:\newline

\noindent
1. -- In all of our models, the disk outflow is capable of driving the shock to breakout from the surface of the star, resulting in an explosion (Figure~\ref{fig:average_shock_radius}). 
While this qualitative result is the same in all our models, the detailed properties of the disk evolution and ejecta depend on the strength of viscous angular momentum transport, on the progenitor star, 
and on the nuclear EOS used in the evolution to BH formation with {\tt GR1D} (Tables~\ref{tab:progenitors} and \ref{tab:supernova}).
\newline

\noindent
2. -- We find that all models produce sufficient $^{56}$Ni to power a type Ic-BL SN light curve.
However, the average asymptotic velocity of the ejecta is too slow, by a factor of $\sim 2-3$
relative to what is needed to account for type Ic-BL SN spectra (Table~\ref{tab:supernova}). The total
kinetic energies of our outflows are in the right range, but our ejecta masses are too high compared to what
is inferred from Ic-BL light curves.
\newline

\noindent
3. -- We find insufficient neutronization of the ejected material to support the production of heavy $r$-process elements (Figure~\ref{fig:all_model_histogram}). While significant neutronization does occur in the disk (Figure~\ref{fig:electron_fraction}), the neutron-rich material is accreted to the BH and not ejected. 
\newline

\noindent
4. -- Neutrino luminosities exhibit a drop of many orders of magnitude at BH formation, followed by a subsequent rise and peak when the disk thermalizes (Figure~\ref{fig:neutrino_luminosity}). 
The duration of the gap in neutrino emission and the magnitude of the peak after disk formation, are dependent on the stellar compactness, accretion rate, and angular momentum profile of the progenitor. This is a diagnostic observable of massive star interiors, should a galactic collapsar occur. 
\newline

\noindent
5. -- In some models the newly formed shocked disk exhibits oscillations during the NDAF phase (Figure~\ref{fig:legendre}). The oscillation frequencies are consistent with characteristic frequencies of the cavity (inverse of sound crossing time and advection time; Figure~\ref{fig:frequency_analysis}). After an oscillatory phase lasting a few seconds, the shock geometry freezes as it begins to expand more rapidly. Generally, the shock waves are extended in one of the polar directions at shock breakout, with the highest viscosity model having a more spherical shape than the others.
\newline

The degree of neutronization of the ejecta depends on the importance of neutrino emission and absorption, which
in turn depends on the thermodynamics of the disk. How close to the BH the disk forms and how
dense it gets depends on the circularization radius (equation~\ref{eqn:rcirc_newtonian}), which in turn depends on the BH mass, the angular momentum profile of the star, and on the accretion rate, which depends on the density profile of the progenitor (or alternatively, the core compactness of the progenitor).
A star with a high core compactness, as well as with density and rotation profiles that decrease slowly with radius would maximize neutronization in the disk. Here we have restricted ourselves to long GRB progenitors that have previously been used in collapsar studies, exploring other progenitors and rotation profiles is left for future work.

Progenitor variation aside, however, we find here that rapid expansion of the shock only begins once the disk has transitioned
to an ADAF phase, due to the decreasing density in the disk, 
which implies that mass ejection is tied to the end of neutronization. 
Ref.~\cite{just_2022} evolves a collapsar disk with no viscosity, finding that while it remains in the NDAF phase 
for its entire evolution and it supports a neutrino-driven wind, it does not eject any significant amounts  of neutron-rich material either.
Thus, \emph{ejection of matter that can support the $r$-process might not be possible if the mass ejection mechanism is thermal} (relying on viscous heating without neutrino cooling, in our case, or on neutrino heating in the inviscid model of \cite{just_2022}). 
Inclusion of magnetohydrodynamics could overcome this hurdle, as material ejected mechanically via Lorentz force from the neutronized disk can bypass the requirement of reaching an ADAF phase for mass ejection (as is 
the case in NS merger disks evolved in MHD, which significantly increase the amount of neutron-rich ejecta
relative to that obtained with viscous hydrodynamics; e.g. \cite{siegel_2018,F19_grmhd,just_2022b,hayashi_2022,curtis_2023}). 
The question of neutrino absorption raising $Y_e$ from its neutronized equilibrium value 
would still remain, however (e.g., \cite{miller_2020}). 
Thus, global, long-term MHD simulations of collapsar disk outflows with good neutrino 
radiation transport are needed to definitively answer the question of whether collapsars can be a relevant $r$-process site.

The entropy per baryon of ejected material spans a broad distribution, with a high entropy tail arising at later times in the simulation reaching several hundred $k_B$ per baryon or more. A
small fraction of the ejecta could therefore (possibly) produce light $r$-process elements in the  high-entropy regime, similar to the conditions in the neutrino-driven winds of some CCSNe models (e.g., \cite{wanajo_2018,witt_2021,wang_2023}). 

The low asymptotic velocities of the ejecta from our models, relative to what is needed to account for the
spectra of type Ic-BL SNe is, like the low degree of neutronization, a consequence of the thermal
nature of mass ejection when using viscous hydrodynamics. In the context of neutron star mergers disk outflows, \cite{FF18} studied the ability of viscous hydrodynamic simulations to produce high-velocity ejecta, over a wide range of (plausible) parameter space, finding that there is a limit to the outflow speed. Subsequent post-merger disk simulations in MHD showed that this limit can easily be overcome by a combination of mechanical ejection by the Lorentz force and neutrino absorption \cite{combi_2023,curtis_2023b,kiuchi_2023,FF23}. We surmise that a similar phenomenon is applicable to collapsar disk outflows, with inclusion of MHD in long-term disk simulations boosting wind speeds to values compatible with observed supernova spectra. 

The usefulness of the gap in neutrino emission between BH formation and collapsar disk formation (Figure~\ref{fig:neutrino_luminosity}) as a diagnostic of supernova physics is contingent on an accurate 
evolution prior to BH formation. In this respect, phenomena such as transient accretion disk formation 
during the protoneutron star phase (e.g., \cite{obergaulinger_2022}) and magnetic effects would alter
the evolution of the neutrino luminosities and cannot be captured by spherically symmetric core-collapse
like we have used here.

The shock oscillations observed during the NDAF phase in some models resemble the standing shock oscillations seen in the post-bounce phase of core-collapse SNe (the `SASI', \cite{blondin_2003,foglizzo_2007}). 
Keeping in mind the qualitative differences between the standing shock in core-collapse SNe and the
shock that bounds the accretion disk in 
collapsars (c.f. Section \ref{s:disk_evolution}),  it is worth noting that in the former,
the oscillation frequencies are tied to oscillations in the neutrino luminosity, which would be observable in a galactic SN \cite{lund_2010,tamborra_2013}, 
as well as to detectable gravitational wave emission (e.g., \cite{kotake_2009,murphy_2009}). 
While our axisymmetric simulations only allow for poloidal oscillations, a three-dimensional model would allow for the
existence of spiral modes. This could in principle result in qualitative differences in the flow dynamics: the 
ADAF phase, during which we find freezing of oscillations, can be unstable to non-axisymmetric perturbations 
\cite{gu_lu_2006,nagakura_yamada_2009}, consistent with the results of \cite{gottlieb_2022}.
A more in-depth analysis of correlations between shock oscillations and temporal fluctuations in the neutrino luminosity will inform the potential for these oscillations to also be an observable of the shocked disk in collapsars. 

An in-depth analysis of nucleosynthesis of the disk outflow, making full use of the 19-isotope network and post-processing of tracer particles, will be presented in a follow up paper.


\appendix

\begin{acknowledgments}
We thank Steven Fahlman, Suhasini Rao, Thierry Foglizzo, and Brian Metzger for helpful discussions.
This research was supported by the Natural Sciences and Engineering Research
Council of Canada (NSERC) through Discovery Grant RGPIN-2022-03463. Support was
also provided by the Alberta Graduate Excellence Scholarship to CD.
The software used in this work was in part developed by the U.S Department of Energy (DOE)
NNSA-ASC OASCR Flash Center at the University of Chicago.
Data visualization was done in part using {\tt VisIt} \cite{VisIt}, which is supported
by DOE with funding from the Advanced Simulation and Computing Program
and the Scientific Discovery through Advanced Computing Program.
This research used computing and storage resources of the
National Energy Research Scientific Computing Center
(NERSC), which is supported by the DOE Office of Science
under Contract No. DE-AC02-05CH11231 (repository m2058).
This research was enabled in part by computing and storage support
provided by Prairies DRI, BC DRI Group, Compute Ontario (computeontario.ca),
Calcul Qu\'ebec (www.calculquebec.ca) and the Digital Research Alliance of Canada (alliancecan.ca). Computations were performed on the Niagara supercomputer at the SciNet HPC Consortium. SciNet is funded by Innovation, Science and Economic Development Canada; the Digital Research Alliance of Canada; the Ontario Research Fund: Research Excellence; and the University of Toronto. 
\end{acknowledgments}

\section{Nuclear Burning and Equation of State}

\subsection{Internal Energy Update \label{sec:internal_energy_update}}

After the hydrodynamic step is complete, the internal energy is first updated by
viscous heating and neutrino heating/cooling from the leakage/absorption scheme
\begin{equation}
\label{eqn:eint_update_neutrinos}
\epsilon^{n+1/2}=\epsilon^n + \left(\frac{1}{\rho\nu}T:T + q_\nu\right) \Delta t\\
\end{equation}
where the superscript denotes time step (all other symbols follow the notation
in Section \ref{sec:flash}).
The subsequent update due to nuclear energy release depends on whether nuclear species are evolved by the nuclear reaction network or 
the NSE solver. 

For $T < 5\times 10^9$\,K, we use the nuclear network to update abundances. The change
in nuclear binding energy is then accounted for in the Newton-Raphson iteration to find the temperature,
instead of its normal direct application as a source term.
\begin{eqnarray}
\label{eqn:abundance_update_network}
&X_i^{n+1}  =  \int_{t^n}^{t^{n+1}}\Theta_i\,dt + \Gamma_{\nu,i}\Delta t\\
\label{eqn:eint_update_network}
&\epsilon^{n+1}(T^{n+1})\big |_{\rho,\mathbf{X}}  =  \epsilon^{n+1/2} + \sum_i B_i \left(X_i^{n+1}-X_i^n \right)\, 
\end{eqnarray}
where $B_i = \chi_i/m_i$ is the nuclear binding energy per unit mass of species $i$,
and the charged-current abundance rate of change terms $\Gamma_{\nu,i}$ are all zero
except for $i=\{{\rm n,p}\}$. The right hand side of equation~(\ref{eqn:eint_update_network}) is then used as the
input internal energy to match with the N-R solver in the Helmholtz EOS.

For $T \geq  5\times 10^9$\,K abundances are determined by the NSE solver for a given $\{\rho,T,Y_e\}$
combination. Instead of equations~(\ref{eqn:abundance_update_network})-(\ref{eqn:eint_update_network}), we have
\begin{eqnarray}
X_{\rm \{n,p\}}^{n+1/2}  =  \Gamma_{\nu,{\rm \{n,p\}}}\Delta t\phantom{HHHHHHHHHHAA\,}\\
Y_e^{n+1/2}  =  \sum_i \frac{Z_i}{A_i} X_i^{n+1/2}\phantom{HHHHHHHHHH}\\
\left[\epsilon^{n+1}(T^{n+1}) - \sum_i B_i X_i^{n+1}(T^{n+1})\right]\bigg |_{\rho,Y_e^{n+1/2}} =\nonumber\\
\label{eqn:eint_update_nse}
\epsilon^{n+1/2} - \sum_i B_i X_i^n
\end{eqnarray}
Equation~(\ref{eqn:eint_update_nse}) defines the new Newton-Raphson function to
obtain the temperature, internal energy, and abundances at step $n+1$ (the NSE
abundances must be updated during each iteration, i.e. it is a nested
Newton-Raphson system). The derivative of this function requires
$(\partial \epsilon/\partial T)_{\rho,Y_e}$, which is computed by the Helmhotz EOS,
and $(\partial X_i/\partial T)_{\rho,Y_e}$, which can be obtained from the
NSE solution at each iteration.

\subsection{Nuclear Statistical Equilibrium (NSE) \label{sec:NSE}}

To obtain the abundances in NSE, we start from the chemical potential for each nuclear species $i$
assuming Maxwell-Boltzmann statistics:
\begin{equation}
\label{eqn:sakur_tetrode}
\mu_i = kT \left[\ln\left(\frac{n_i}{n_{{\rm Q},i}} \right) - \ln\omega_i\right] - \chi_i
\end{equation}
where $n_i$ is the number density, $\omega_i$ is the partition function, $\chi_i$ is the
nuclear binding energy, and
\begin{equation}
n_{{\rm Q},i} = \left(\frac{m_i kT}{2\pi \hslash^2} \right)^{3/2}.
\end{equation}
is the quantum concentration (e.g., \cite{kittel_1980}). 
Solving for the number density in
equation (\ref{eqn:sakur_tetrode}) and expressing as a mass fraction yields
\begin{equation}
\label{eqn:nse_mass_fraction}
X_i = \frac{m_i}{\rho}\omega_i\,n_{{\rm Q},i}(T)\, \exp\left(\frac{\mu_i + \chi_i}{kT} \right).
\end{equation}
Nuclear statistical equilibrium is obtained by imposing chemical equilibrium for each
species
\begin{equation}
\label{eqn:chemical_equilibrium}
\mu_i = N_i \mu_{\rm n} + Z_i \mu_{\rm p},
\end{equation}
where $N_i = A_i - Z_i$ is the number of neutrons in each nucleus, as well as mass
and charge conservation
\begin{eqnarray}
\label{eqn:mass_cons_nse}
\sum_i X_i & = & 1\\
\label{eqn:charge_cons_nse}
\sum_i \frac{Z_i}{A_i}X_i & = & Y_e.
\end{eqnarray}
In practice, calculation involves doing a non-linear root
find\footnote{We use the NSE solver written by F. Timmes, available at {\tt cococubed.asu.edu}}
for $\{\mu_{\rm n},\mu_{\rm p}\}$ by
replacing equations~(\ref{eqn:nse_mass_fraction})-(\ref{eqn:chemical_equilibrium}) into
(\ref{eqn:mass_cons_nse})-(\ref{eqn:charge_cons_nse}),
for given values of $\{\rho, T, Y_e\}$.

The temperature derivatives of the abundances in NSE can be obtained by
replacing equation~(\ref{eqn:chemical_equilibrium})
in equation~(\ref{eqn:nse_mass_fraction}) and differentiating
\begin{eqnarray}
\left(\frac{\partial X_i}{\partial T}\right)_{\rho,Y_e} & = &
 \frac{X_i}{T}\left[\frac{3}{2}
     + \frac{N_i}{k}\left(\frac{\partial\mu_{\rm n}}{\partial T}\right)_{\rho, Y_e}
     +\frac{Z_i}{k}\left(\frac{\partial\mu_{\rm p}}{\partial T}\right)_{\rho,Y_e} \right.\nonumber\\
\label{eqn:dXdT_nse}
 && \left. - \frac{1}{kT}\left(N_i\mu_{\rm n}+Z_i\mu_{\rm p}+\chi_i\right)\right],
\end{eqnarray}
where we have assumed that the partition function $\omega_i$ is constant; inclusion
of that term (if known) is straightforward.
The derivatives of the chemical potentials can be obtained by differentiating
equations~(\ref{eqn:mass_cons_nse})-(\ref{eqn:charge_cons_nse}) with respect to
temperature, and substituting equation~(\ref{eqn:dXdT_nse}), which yields a $2\times 2$ linear
system that can be solved analytically once $\{\mu_{\rm n},\mu_{\rm p}\}$ are known:

\begin{eqnarray}
&\left(\sum_i N_i X_i \right)\left(\frac{\partial \mu_{\rm n}}{\partial T}\right)_{\rho, Y_e} + 
\left(\sum_i Z_i X_i \right)\left(\frac{\partial \mu_{\rm p}}{\partial T}\right)_{\rho,Y_e} =\nonumber \\
&\left[\frac{1}{T}\sum_i X_i \left(N_i\mu_{\rm n}+Z_i\mu_{\rm p}+\chi_i\right)-\frac{3}{2}k\right]\\
&\left(\sum_i \frac{Z_i N_i}{A_i} X_i \right)\left(\frac{\partial\mu_{\rm n}}{\partial T}\right)_{\rho, Y_e} +
\left(\sum_i \frac{Z_i^2}{A_i} X_i \right)\left(\frac{\partial\mu_{\rm p}}{\partial T}\right)_{\rho,Y_e} =\nonumber \\
&\left[\frac{1}{T}\sum_i \frac{Z_i X_i}{A_i} \left(N_i\mu_{\rm n}+Z_i\mu_{\rm p}+\chi_i\right)-\frac{3}{2}k Y_e\right].
\end{eqnarray}

\section{Variable Floors}
\label{sec:variable_floors}

\subsection{Density, pressure and internal energy floors}

We use variable floors with radial and polar angle dependencies for  
density, pressure, and internal energy. The general functional form is
\begin{equation}
W_{\rm floor}(r,\theta) = W_0 \cdot f_{\rm floor}(r,\theta)
\end{equation}
where $W$ stands for any of $\{\rho,p,\varepsilon\}$, $W_0$ is a constant value, and $f_{\rm floor}$ is 
a dimensionless function with a maximum of 1 which contains the radial and polar angle dependencies. 
The floor function is in turn a product of radial and angular factors:
\begin{equation}
f_{\rm floor}(r,\theta) = f_r(r)\cdot f_\theta(\theta) 
\end{equation} 
The radial factor is a 5-piece power-law function given by:
\begin{equation}
\begin{aligned}
f_r(r) =  
\begin{cases}
1 & r<R_{1} \\
\left(\frac{R_{1}}{r}\right)^{s_{1}} & R_{1} < r < R_{2} \\
\left(\frac{R_{1}}{R_{2}}\right)^{s_{1}}\left(
\frac{R_{2}}{r}\right)^{s_{2}} & R_{2} < r < R_{3} \\
\left(\frac{R_{1}}{R_{2}}\right)^{s_{1}}\left(
\frac{R_{2}}{R_{3}}\right)^{s_{2}}\left(\frac{R_{3}}{r}\right)^{s_{3}}
& R_{3} < r < R_{4} \\
\left(\frac{R_{1}}{R_{2}}\right)^{s_{1}}\left(
\frac{R_{2}}{R_{3}}\right)^{s_{2}}\left(\frac{R_{3}}{R_{4}}\right)^{s_{3}}\left(\frac{R_{4}}{r}\right)^{s_{4}}
& r > R_{4} \\
\end{cases}
\end{aligned}
\end{equation}
where $R_{i}$ and $s_{i}$ are constant transition radii and slopes, respectively.
This functional form is chosen to approximately follow the radial stellar profile,
with normalization values $W_0$ such that each floor stays a few orders of magnitude below
the actual hydrodynamic variable throughout the simulation.
Transition radii and slopes were determined through comparison to the initial stellar profile, as well as iterative
analysis of initial model evolution, with slopes ranging from $0.5-50$.
The normalization coefficients are model-dependent, falling in the range $\rho_0 = 10^4-10^5\,$g$\,$cm$^{-3}$ for
density, $P_0 = 10^{22}-10^{23}\,$dyn$\,$cm$^{-2}$ for pressure, and $\epsilon_0 = 10^{17}-10^{18}\,$erg\,g$^{-1}$
for internal energy.

The angular factor is:
\begin{equation}
f_{\theta}(\theta) =
(1-\tilde{\theta}_{\rm eq})\cos(\theta)^{2\tilde{\theta}_w}+\tilde{\theta}_{\rm eq}
\end{equation}
where $\tilde{\theta}_{\rm eq}$ is the equatorial floor, and
$\tilde{\theta}_w$ is a width factor that controls how quickly the floor drops off away from the poles toward the equator.
This functional form is used to deal with the low density funnel near the poles,
without interfering with the disk at the equator. 

\subsection{Temperature floor}

Our temperature floor is given by:
\begin{equation}
T_{\rm floor}(r) =
\rm{max}\left[10^7\,\rm{K}\cdot\left(\frac{50\,\rm{km}}{r}\right)^2,10^4\,\rm{K}\right],
\end{equation}
where the minimum value is associated with the bottom of the Helmholtz EOS table in {\tt FLASH}.
This functional form is necessary to deal with
problematic cells at the shear interface between the 
shocked disk and the low-density funnel near the inner radial boundary.   

\bibliographystyle{apsrev4-2}
\bibliography{ms}

\end{document}